\documentclass[preprint,3p,times]{elsarticle}

\graphicspath{ {./elsarticle/} }

\usepackage{elsarticle/ecrc}

\volume{00}

\firstpage{1}

\journalname{Journal of Computational Science}

\runauth{D. Bartuschat and U. R{\"u}de}

\jid{JCompSci} 

\jnltitlelogo{J. Comput. Sci.}

\usepackage{lmodern}
\usepackage{amsmath}
\usepackage{array}
\usepackage[hang]{subfigure} 

\usepackage{units}
\usepackage[mode=text,exponent-product = \cdot]{siunitx} 
\usepackage{multirow}
\usepackage{booktabs}   

\usepackage[utf8]{inputenc}
\usepackage{amsmath}
\usepackage{amsfonts}
\usepackage{amssymb}

\usepackage{fontenc}
\usepackage[hidelinks]{hyperref}
\usepackage[textwidth=\columnwidth]{todonotes}

\usepackage{booktabs}

\usepackage[ruled,vlined]{algorithm2e}
\usepackage{algorithmic}  
\usepackage{color}

\usepackage{caption} 

\usepackage{listings}
\newcommand{\lstverb}{\lstinline[language=C++,%
   basicstyle=\small\ttfamily,%
   breaklines=true,%
   keywordstyle=\bfseries\color{red}%
   ]}

\newcommand {\lapOp} {\, \Delta \,} 
\newcommand {\gradOp} {\nabla } 
\newcommand {\divOp} {\nabla \cdotp} 

\newcommand{\partderivT}[1]{ 
  \dfrac{\partial #1}{\partial t}
}

\newcommand\restr[2]{{
  \left.\kern-\nulldelimiterspace 
  #1 
  \vphantom{\big|} 
  \right|_{#2} 
}}

\newcommand{\eg}{\mbox{e.\,g.}\xspace}
\newcommand{\ie}{\mbox{i.\,e.}\xspace}
\newcommand{\cf}{\mbox{cf.}\xspace}
\newcommand{\Fig}[1]{\mbox{Fig.\,\ref{#1}}}
\newcommand{\Sect}[1]{\mbox{Sec.\,\ref{#1}}}
\newcommand{\Tab}[1]{\mbox{Tab.\,\ref{#1}}}
\newcommand{\Eqn}[1]{\mbox{Eqn.\,(\ref{#1})}}
\newcommand{\Alg}[1]{\mbox{Alg.\,\ref{#1}}}

\newcommand{\walberla}{\textsc{waLBerla}\xspace}
\newcommand{\Walberla}{\textsc{WaLBerla}\xspace}

\DeclareMathAlphabet{\mathpzc}{OT1}{pzc}{m}{it}
\newcommand{\pe}{$\mathpzc{pe}$\xspace}

\begin{document}

\begin{frontmatter}

\dochead{}

\title{A Scalable Multiphysics Algorithm for Massively Parallel Direct Numerical Simulations of Electrophoresis}

\author[LSS]{Dominik Bartuschat\corref{cor1}}
\ead{dominik.bartuschat@cs.fau.de}
\cortext[cor1]{Corresponding author}
\author[LSS,CERFACS]{Ulrich R\"ude}
\address[LSS]{Lehrstuhl f\"ur Systemsimulation, Friedrich-Alexander Universit\"at Erlangen-N\"urnberg, Cauerstrasse 11, 91058 Erlangen, Germany}
\address[CERFACS]{Parallel Algorithms Group, CERFACS, 42 Avenue Gaspard Coriolis, 31057 Toulouse, France}

\begin{abstract}
In this article we introduce a novel coupled algorithm for massively parallel direct numerical
simulations of electrophoresis in microfluidic flows. This multiphysics algorithm
employs an Eulerian description of fluid and ions, combined with a Lagrangian
representation of moving charged particles. The fixed grid facilitates
efficient solvers and the employed lattice Boltzmann method can efficiently 
handle complex geometries. Validation experiments with more than $70\,000$ time steps are presented, together with
scaling experiments with over ${4\cdot10^{6}}$ particles and ${1.96\cdot10^{11}}$ grid cells
for both hydrodynamics and electric potential. 
We achieve excellent performance and scaling on up to $65\,536$ cores of a current supercomputer.
\end{abstract}


\end{frontmatter}




\section{Introduction}

\subsection{Motivation}
The motion of charged particles in fluids under the influence of electric fields occurs in a wide range of industrial, medical, and biological processes.
When the charged particles are immersed in liquids, their migration caused by electric fields is termed electrophoresis.
Due to the complex interplay of the physical effects involved in such particle-laden electrokinetic flows,
numerical simulations are required to analyze, predict, and optimize the behavior of these processes.
To this end, we present a parallel multiphysics algorithm for direct numerical simulations of electrophoretic particle motion.

Industrial applications that involve electrophoretic effects are electrofiltration~\cite{Ptasinski1992ElFieldDrivenSepar,Zhang_2000_Electrofiltration,Weng2006EMF}
and electro-dewatering~\cite{Mahmoud2010_Dewatering}. 
Moreover, electrophoresis is utilized in electrophoretic deposition techniques for
fabricating advanced materials~\cite{Besra2007ReviewEPD} 
and especially ceramic coatings~\cite{Zhitomirsky2002EPDCeramic,Sarkar2012EPD} in material science.
Electrophoresis and electric fields are also applied in many medical and biological applications.
The trend towards miniaturization of analysis processes has lead to the development of micro total analysis systems.
Due to their high portability, reduced costs, fast operation, and high sensitivity~\cite{KangLi:2009:EKFParticlesCells,Bhagat2010CellSepar},
the design of such lab-on-a-chip systems has been a highly active area of research for many years.
These microfluidic systems require only small samples of liquid and particles,
which are transported, manipulated, and analyzed in structures of length scales from several \si{\nano\meter} to \SI{100}{\micro\meter}.
Therefore, microfluidic separation and sorting of particles and cells are important steps of diagnostics in such systems~\cite{gossett2010label,pamme2007ContinFlowSep}.
Many of the employed techniques utilize electric fields to
manipulate, separate, and sort biological particles and macromolecules~\cite{KangLi:2009:EKFParticlesCells},
such as cells~\cite{Bhagat2010CellSepar,gossett2010label} or DNA~\cite{Hert2008DNAsequencingELPS}. 

At the small scales of microfluidic analysis systems, flow measurements are difficult or even impossible.
Moreover, the complex coupling of hydrodynamic and electrostatic effects involved in electrophoretic processes
make predictions of electrophoretic motion challenging, especially for large numbers of particles.
Therefore, numerical simulations are essential to aid the design and optimization of electrophoretic systems.
The different physical effects in electrophoretic deposition can be better understood from insight gained in simulations.
By means of such simulations, electrophoretic sorting in lab-on-a-chip systems can be optimized for maximal throughput, sorting efficiency, and sorting resolution.
A review of simulation methods for electrophoretic separation of macromolecules is given in \cite{Slater2008ModelingSeparMacromol}.
Also industrial applications of electrophoretic deposition can be optimized with the help of simulations, as presented in \cite{Keller2015ElphorDeposSim} for a coating process.

\subsection{Multiphysics Coupling Strategy}
\begin{figure}[h]
  \centering
     \includegraphics[width=0.68\textwidth]{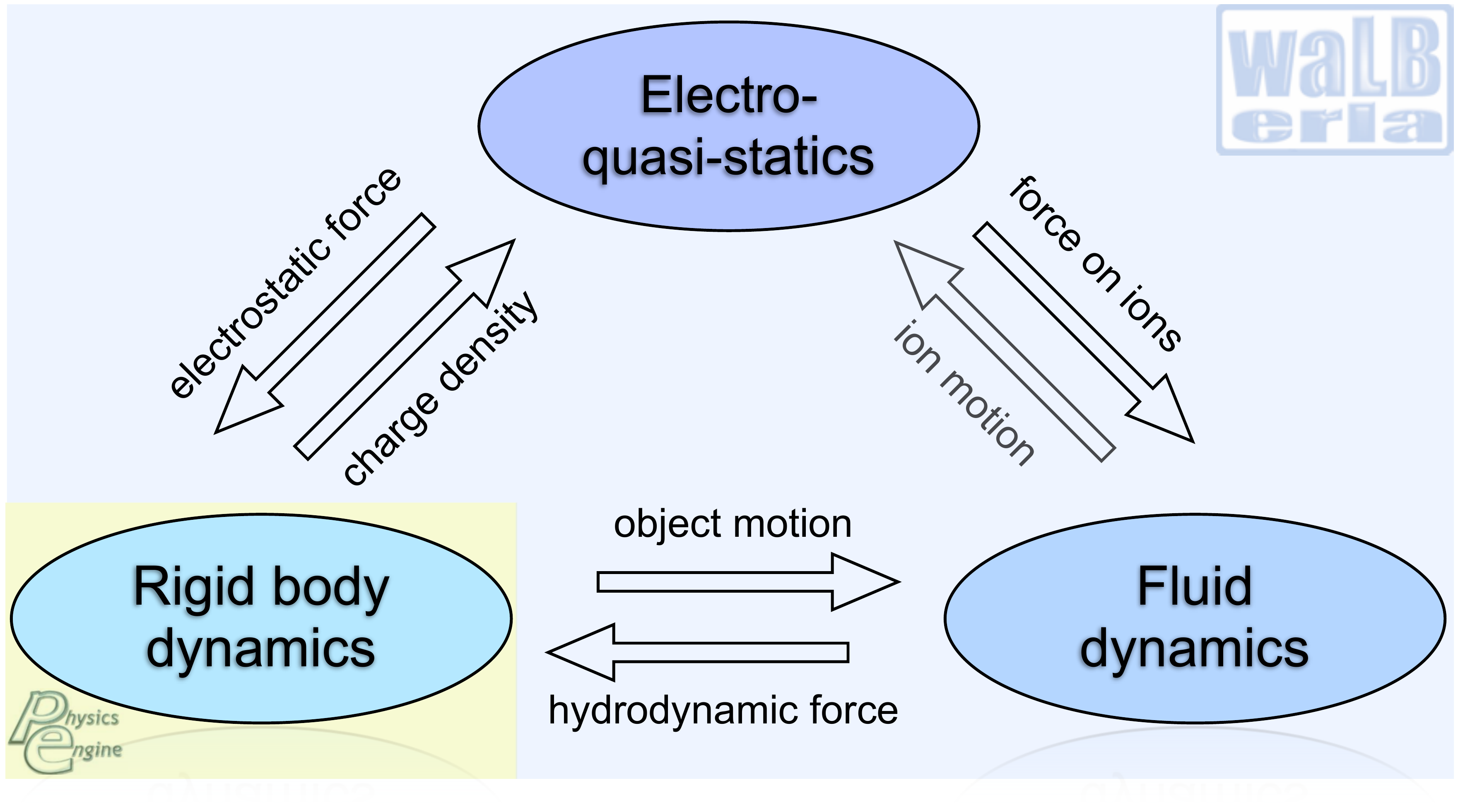}
    \caption{Coupled physical effects of electrophoresis simulated with~\walberla and the physics engine \pe.\label{fig:EKF_Multiphysics_Interactions}}
\end{figure}

For simulations of electrokinetic flows with electrophoretic particle motion, the coupling between three system components must be modeled:
charged objects, fluid flow, and electric effects.
The interacting physical effects are sketched in \Fig{fig:EKF_Multiphysics_Interactions}.
In the simulation method introduced in this article, the motion of the rigid, charged particles is modeled with Newtonian mechanics.
The motion of the surrounding fluid, which exerts hydrodynamic forces on the particles, is described by the incompressible Navier-Stokes equation.
To capture fluid-particle interactions, the hydrodynamic forces and the influence of the particle motion on the fluid are modeled, based on the momentum exchange between fluid and particles.
In this way, long-range hydrodynamic interactions between individual particles and between particles and walls are recovered.
~\\
Moreover, electrostatic forces exerted by applied electric fields on the charged particles are modeled, which are the main cause of electrophoretic motion.
The varying positions of the charged particles in return affect the electric potential distribution in the simulation domain, based on their surface charge.
Such a charge is carried by most biomolecules such as cells, proteins, and DNA~\cite{sheehan2013physical}.
In fact, most substances acquire surface charges when they get into contact with an aqueous medium~\cite{probstein1989Physicochem} or electrolyte solution~\cite{chang2009electrokinetically}. \\
Electrostatic forces additionally act on the ions in the fluid and affect the fluid motion via body forces present in locations of net charge.
This net charge originates from the repulsion of co-ions in the fluid of the same polarity as the surface charge and from the attraction of counter-ions.
The ions in the fluid are transported with the fluid flow, which in turn alters the electric potential distribution.
Ion motion in the fluid due to an electric field is governed by the Nernst-Planck equation,
an advection-diffusion equation that employs a continuum description and treats the ions as point charges.
For flows in which diffusion strongly dominates over advection and therefore quasi-thermodynamic equilibrium can be assumed to hold,
as considered in this paper, the electric potential due to the ion charge distribution is governed by the Poisson-Boltzmann equation.

The region of the particle's surface charge and of excess counter-ions in the fluid is denoted as \emph{electric double layer} (EDL).
According to the Stern model~\cite{Stern1924EDL} employed in this article, this double layer comprises a region of ions attached to the surface 
and a diffuse part in which the ion concentration follows a Boltzmann distribution.
At the surface of shear between the particle and the surrounding diffuse double layer, the characteristic $\zeta$-potential is defined.
The employed equilibrium considerations based on the Poisson-Boltzmann equation capture the dominant retardation effect of electrophoretic motion.
This effect describes the retardation of the charged particle motion by the action of the applied electric field on the opposite net charge in the surrounding EDL.
At high $\zeta$-potentials, additionally the weaker relaxation effect occurs~\cite{probstein1989Physicochem} 
that is caused by a distortion of the EDL and can be captured by the Nernst-Planck equation.

In this article, we present an efficient parallel multiphysics simulation algorithm for electrophoresis on a fixed Eulerian grid with a Lagrangian representation of moving particles.
The particles are represented by the physics engine \pe{}~\cite{preclik2015ultrascale,Iglberger:2009:CSRD} as geometrically fully resolved three-dimensional objects.
Dependent on the electrostatic and hydrodynamic forces acting on the particles, the \pe\ computes their trajectories by rigid body dynamics and additionally resolves particle collisions.
The \pe is coupled to \walberla \cite{Feichtinger2011105,godenschwager2013framework,schornbaum2016massively},
a massively parallel simulation framework for fluid flow applications that employ the lattice Boltzmann method (LBM)~\cite{Chen:98:LBM_For_Fluid_Flows,aidun2010lattice}.
By means of a modular software structure that avoids dependencies between modules, functionality from different modules can be combined flexibly.
To model fluid-particle interactions, the LBM momentum exchange method~\cite{Ladd_1993_PartSuspPt1,nguyen2002lubrication} implemented in \walberla{}~\cite{Goetz:2010:SC10} is applied.
For the electrophoresis simulations, the LBM is performed with the two-relaxation time collision operator~\cite{ginzburg2004lattice,ginzburg2008two} and an appropriate forcing term for the electric body force due to the ions in the fluid.
The electric potential is represented by the Debye-H{\"u}ckel approximation of the Poisson-Boltzmann equation that is discretized with finite volumes whose mesh structure naturally conforms to the lattice Boltzmann (LB) grid.
This discretization also facilitates accommodating variable and discontinuous dielectricity values that vary in time according to the particle positions, as required for simulating dielectrophoretic effects.
By means of the \walberla solver module introduced in \cite{Bartuschat:2014:CP} together with the cell-centered multigrid solver implemented therein,
the resulting linear system of equations is solved.
Since the counter-ions lead to a quicker decay of the electric potential compared to the long-range electric potentials modeled in \cite{Bartuschat:2014:CP},
the parallel successive over-relaxation (SOR) method implemented in this module is an adequate choice.

In a previous article, we have shown that the implemented fluid-particle interaction algorithm for arbitrarily shaped objects
efficiently recovers hydrodynamic interactions also for elongated particles \cite{Bartuschat:2014:Tumbling}.
Moreover, we presented a parallel multiphysics algorithm for charged particles in the absence of ions in the fluid in \cite{Bartuschat:2014:CP}.
We have therein shown that several millions of charged particles with long-range hydrodynamic and electrostatic interactions can be simulated with excellent parallel performance on a supercomputer.
The present paper extends these simulation algorithms by also considering ions in the EDL around the particles and their effect on the fluid motion,
and presenting suitable parallel coupling techniques. 
Together with the full four-way coupling of the fluid-particle interaction \cite{Bartuschat:2014:CP}, the coupling with the quasi-equilibrium representation of the electric effects results in a 7.5-way interaction.

\subsection{Related Work}
In the following, we give an overview of numerical methods for simulations of electrophoretic phenomena that have been developed for different resolution levels.
At the coarsest modeling scale, both fluid and solid phase are described by an Eulerian approach.
These continuum models represent the charged species in terms of concentrations and
are well suited for simulations with large numbers of unresolved particles.
In \cite{Jomeh2012Electrophresis} two-dimensional electrophoresis simulations of biomolecule concentrations are presented.
These finite element simulations consider the effect of reactive surfaces and the electric double layer is included through the biomolecule charge.
Also three-dimensional parallel simulations of electrophoretic separation with continuum approaches have been reported.
The finite element simulations in \cite{kler2011modeling} consider the buffer composition and the $\zeta$-potential at channel walls.
In \cite{Chau2013Asynchron3Delectrophor,Chau2007ContinFlowElectrophor} mixed finite element and finite difference simulations of protein separation were performed on up to 32 processes.

At the finest level of resolution, fluid, ions, and particles are simulated by Lagrangian approaches.
These explicit solvent methods~\cite{giera2015mesoscale} typically apply coarse-grained molecular dynamics (MD) models to
describe the motion of fluid molecules and incorporate Brownian motion~\cite{Sman2009ConfSuspFlow}.
The mesoscale \emph{dissipative particle dynamics}
method is applied in \cite{Smiatek2011DPDeofEP} to simulate electrophoresis of a polyelectrolyte in a nanochannel.
Another explicit solvent method is presented in \cite{Wang2009DNAelphorCGMD} for simulating DNA electrophoresis, modeling DNA as a polymer.
In both methods the polymer is represented by bead-spring chains with beads represented by a truncated Lennard-Jones potential and connected by elastic spring potentials.
Explicit solvent models, however, are computationally very expensive, especially for large numbers of fluid molecules due to pairwise interactions \cite{Wang2009DNAelphorCGMD}.
Moreover, the resolution of solvent, macromolecules, and ions on the same scale limits the maximal problem sizes that can be simulated \cite{kuron2016moving}.
Also the mapping of measurable properties from colloidal suspensions to these particle-based methods is problematic \cite{giera2015mesoscale}.
~\\
The high computational effort is significantly reduced in implicit particle-based methods that incorporate hydrodynamic interactions into the inter-particle forces.
Such methods are applied in \cite{giera2015mesoscale} and \cite{park2012direct} to simulate electrophoretic deposition under consideration of Brownian motion and van der Waals forces.
Nevertheless, these methods are restricted to few particle shapes and hydrodynamic interactions in Stokes flow.

Euler-Lagrange methods constitute the intermediate level of resolution.
These approaches employ Eulerian methods to simulate the fluid phase,
whereas the motion of individual particles is described by Newtonian mechanics.
For simulations of particles in steady-state motion, the resolved particles can be modeled as fixed while the moving fluid is modeled by an Eulerian approach.
In \cite{Hsu2011ElphorToroid} the finite volume method is applied to simulate electrophoresis of up to two stagnant toroids in a moving fluid,
employing the H{\"u}ckel approximation for a fixed $\zeta$-potential and different electrical double layer thicknesses.
The steady-state electrophoretic motion of particles with low surface potentials under a weak applied electric field in a charged cylindrical pore
is simulated in \cite{Hsu2005BoundaryEffectElectrophorCylinder} for a single cylinder and in \cite{Hsu2007Electrophoresis} for two identical spheres.
In both cases, a two-dimensional simulation with a finite element method for Stokes flow and H{\"u}ckel approximation is performed, exploiting the axial symmetry of the problem.\\
For electrophoresis at steady state \emph{perturbation approaches} can be employed that are based on the assumption that the double layer is only slightly distorted from the equilibrium distribution
for weak applied electric fields (w.r.t.\ the field in the EDL).
In addition to the equilibrium description based on the Poisson-Boltzmann equation,
small perturbations in the equilibrium EDL are considered in terms of
linear correction terms in the applied electric field for the ion distribution and the electric potential (see \eg{} \cite{Masliyah:2006:ElectrokinColTransp}).
Using a perturbation approach with finite elements,
the electrophoresis of two identical spheres along the symmetry axis of a cylindrical domain at pseudo steady-state were studied in \cite{Tseng20151ElectophorSpheres}.
Additionally to the hydrodynamic and electric interactions, these axisymmetric simulations consider van der Waals forces for particles in close proximity.
In \cite{Schmitz2013ElkinCollSys} a perturbation approach is applied to simulate a single colloid in a rest frame with periodic boundary conditions (BCs).
The zeroth-order perturbation corresponds to the Poisson-Boltzmann equation, which is solved by a constrained variational approach suggested in \cite{Baptista2009PBEsolver}.
For the first-order perturbation, the stationary Stokes equation is solved by a surface element method,
the convection-diffusion for ionic concentrations by a finite volume solver, 
and the Poisson equation by a fast Fourier Transform method.

More sophisticated Euler-Lagrange methods include direct numerical simulation (DNS) models
that represent the moving particles as geometrically fully resolved objects.
These methods for particulate flows include approaches with \emph{body-fitted moving meshes} and \emph{fixed meshes}.
Moving meshes can be represented by the Arbitrary-Lagrangian-Eulerian (ALE) formulation~\cite{Hu2001ALE,Hughes1981LEFE}
that employs moving unstructured meshes for fluid-particle interaction problems.
Such an ALE method is applied in \cite{Ye2004TransElphorMotionSpher} to simulate electrophoresis of a single particle surrounded by a thin electrical double layer.
The moving-mesh techniques require re-meshing when the distortion of the updated mesh becomes too high,
and subsequent projection of the solution onto the new mesh. 
This overhead is circumvented in \emph{fixed-mesh techniques} that allow the use of regular grids and therefore the application of efficient solvers.
The \emph{fluid particle dynamics} method~\cite{Tanaka2000FPDintro} falls into the latter category, 
solving the Navier-Stokes and continuity equation on a fixed lattice, and representing the moving solid particles by fluid particles of very high viscosity.
By means of a concentration field that represents the particle distribution, the particles affect the fluid viscosity and, together with forces acting on the particles, the body force term of the Navier-Stokes equation.
The rigid particles are geometrically modeled by the Lennard-Jones potential and their motion is described by Newtonian mechanics~\cite{Tanaka2000FPDintro}. 
This technique is applied to simulate electrophoretic deposition of two charged particles in~\cite{Kodama2004FPDelectrophor} and electrophoretic separation in~\cite{Araki2008FPDelphorSepar}.
In these simulations the electrostatic interactions of the particles are modeled in terms of the body force field,
together with the advection and diffusion of ions and the resulting effect of the applied electric field on the fluid motion.
With this method, however, the particle rigidity is imposed by very high viscosity values that restrict the time step size~\cite{Nakayama2005SPMintro}
and the Lennard-Jones potential restricts the particles shapes to spheres.
The \emph{smoothed profile method} (SPM)~\cite{Nakayama2005SPMintro} circumvents this time-step constraint by directly modeling the particles as solid objects.
Inside the particles and at solid-fluid boundaries that are represented by diffuse interfaces, a body force is imposed on the fluid to model the effect of the particle motion on the fluid.
The fluid is again modeled on a fixed Cartesian grid and the particle motion with Newtonian mechanics, where particle overlaps are typically prevented by a truncated Lennard-Jones potential.
With this method, electrophoresis of charged spherical particles is simulated in \cite{Kim2006DNSelectrophorSPM} for a constant, uniform electric field and in~\cite{Shih2014ElphoresisSPM} for an oscillating electric field.
In both articles, the ion number concentration is modeled by an advection-diffusion equation to recover non-equilibrium double layer effects.
The SPM is also applied in \cite{Luo2010SPMelectrophoresis} to simulate electrophoresis of single cylinders and microtubules, employing the equilibrium representation of the EDL. 
A further fixed-mesh technique is the \emph{immersed boundary method}~\cite{Peskin1972FlowPatAroundHrt,mittal2005immersed,Uhlmann2005DFM}, where the rigid body motion is imposed on the flow by body forces applied at the particle boundaries.
This method, combined with a finite volume method for solving the steady-state Poisson-Nernst-Planck equation system, is applied in \cite{Kang2013ElphorMultiple} to simulate the electrophoretic motion of up to three spherical particles in a two-dimensional setup.

Lattice-Boltzmann based methods are very well suited for parallel direct numerical simulations of fluid-particle interactions on fixed Cartesian grids.
Both the Lagrangian particles and ions are often explicitly modeled by molecular dynamics approaches that represent the rigid objects by repulsive potentials.
In \cite{Lobaskin2004ElectrophorLangevin} the electrophoresis of a colloidal sphere immersed in a fluid with counter-ions is simulated, modeling the solvent by a lattice Boltzmann method. 
The charged sphere modeled with molecular dynamics is represented by a raspberry model that comprised several beads connected by the finitely extensible nonlinear-elastic (FENE) potential.
Using a modified raspberry model with two spherical shells of beads solidly attached to a larger spherical particle, 
this method is extended in \cite{Molotilin2016electrophoresis} to simulate the electrophoresis of a spherical Janus particle.
The partially uncharged particle is surrounded by anions and cations represented by charged beads.
Electrophoresis simulations for a single highly charged spherical macro-ion in an electrolyte solution with explicitly modeled positive and negative micro-ions are presented in~\cite{chatterji2010role}.
Since the coupling of fluid and macro-ion is performed via several particle boundary points, a single spherical particle is sufficient to represent the macro-ion.
A similar LB-MD method is applied in \cite{Hickey2014LBelphorStreching} to simulate the stretching of a charged polyelectrolyte between parallel plates.
The polyelectrolyte immersed in a liquid with explicitly modeled counter- and co-ions is modeled by beads bonded together by the FENE potential.
In all these LB-MD simulations with explicitly modeled ions, hydrodynamic interactions are simulated with the LBM and thermal fluctuations are added to both the fluid and the MD objects.
The high computational effort for modeling each individual ion by means of molecular dynamics, however, restricts the maximum feasible problem size.
With these approaches, only a limited number of ions per colloidal particle can be simulated and the colloid radius is typically restricted to one order of magnitude larger than the ion size~\cite{Schmitz2013ElkinCollSys}.
Therefore, approaches based on continuum descriptions of the suspended ions are more practical for simulations of many or for larger charged particles.

Alternatively to the continuum approach based on the Poisson-Boltzmann equation employed in this article,
electrophoresis can be simulated with the \emph{link-flux method}~\cite{capuani2004discrete}
that models the advection and diffusion of ions in terms of Nernst-Planck equations.
The link-flux method employs the LBM for fluid dynamics and models ion motion in terms of fluxes between lattice cells.
In~\cite{giupponi2011determination} this method is compared to a LB-Poisson-Boltzmann approach for a fixed spherical particle in a periodic three-dimensional domain.
The aim is to examine the influence of particle motion and counter-ion concentration on the $\zeta$-potential,
leading to the conclusion that for weakly perturbing electric fields or low P{\'{e}}clet numbers 
the equilibrium and dynamic $\zeta$-potentials are indistinguishable.
The link-flux method is extended in~\cite{kuron2016moving} to support moving particles in combination with the LB momentum exchange method.
To ensure charge conservation in electrophoresis simulations, appropriate moving BCs for the solute fluxes are introduced.
This method is verified in~\cite{kuron2016moving} by electrophoresis simulations of up to eight particles.

\subsection{Objectives and Outline}
The primary goal of this paper is the introduction of a parallel multiphysics algorithm for electrophoresis simulations
together with validations of the physical correctness of the coupled algorithm for different particle sizes.
For this algorithm, \walberla{} is augmented by an efficient boundary handling method that is able to treat electric potential BCs on the moving particles.
Moreover, a joint parameterization for the different coupled numerical methods is introduced.
To achieve excellent computational performance, a matrix-free representation of the linear system based on a {\em stencil} paradigm is used in the solver module~\cite{Bartuschat:2014:CP}.
For the linear Debye-H{\"u}ckel approximation it is systematically exploited that these stencils are almost uniformly identical throughout the simulation domain.
The validation runs were performed on up to \num{8192} parallel processes of a modern supercomputer.
Moreover, simulation results for the electrophoretic motion of a single particle in a micro-channel are presented,
including visualizations of the electric potential distribution and of the resulting flow field around the particle.
Finally, we present performance results on a modern supercomputer. Very good scaling behavior is shown for more than \num{196e9} lattice cells on up to \num{65 536} cores. 
In this case, more than four million fully resolved charged objects and their surrounding double layers are simulated, together with the interactions with the fluid.
To the best of our knowledge, electrophoresis simulations of that size are unprecedented.

The equilibrium considerations in the present paper recover the predominant retardation effect due to an opposing electrostatic force on the net opposite charge in the electrical double layer that counteracts the particle motion.
For the presented method, a computationally cheap and flexible SOR method is sufficient to solve the electric potential equations.
With our approach we aim for simulations of millions of charged particles as in~\cite{Bartuschat:2014:CP}.
For these large numbers of particles the dynamics of an electrical double layer as in~\cite{kuron2016moving} is computationally too expensive, even on modern supercomputers.

This paper is structured as follows:~The physical background of fluid-particle interactions and electrophoresis are described in \Sect{Sec:FPI} and \Sect{Sec:ElkinFlow}, respectively.
In \Sect{Sec:NumModeling}, the employed LB-momentum exchange method for fluid-particle interactions is outlined, 
together with the finite volume discretization and the common parameterization concept for the coupled multiphysics methods.
Then the extension of the \walberla\ framework for the electrophoresis algorithm is described in \Sect{Sec:ExtWalberlaElph}.
Finally validation results for the electrophoretic motion of a spherical particle and visualizations of the resulting flow field and electric potential distribution are presented in \Sect{Sec:ElPhorResults}
before conclusions are drawn in \Sect{Sec:Conclusion}.

\section{Fluid-Particle Interaction \label{Sec:FPI}}

The macroscopic description of fluid behavior is
based on the continuum hypothesis (\cf Batchelor~\cite{batchelor1979introduction})
that allows to consider a fluid as continuum.
Fluid properties are then represented by the macroscopic quantities
density $\rho_f$, velocity $\vec{u}$, and pressure $p$, as functions of space and time.
In terms of these quantities, fluid dynamics is described by conservation laws for mass, momentum, and energy.
In this article isothermal flows are considered, and therefore the energy equation does not have to be solved.
Moreover, non-continuum effects that become relevant for gas flows at very small scales \cite{vanderHoef2006Multiscale} are assumed to be negligible.
Therefore, no-slip BCs are applied at solid-fluid interfaces,
and slip velocities due to non-continuum Knudsen layer effects are not considered. 

Conservation of mass is described by the continuity equation.
For incompressible fluids the density
is not affected by pressure changes~\cite{batchelor1979introduction,LandauLifschitzEngl} 
and the continuity equation reads
\begin{equation}
   \divOp \vec{u} = 0.
   \label{Eq:IncompContEq}
\end{equation}

Conservation of momentum in a viscous, compressible fluid can be described in terms of the momentum flux density tensor $\underline{\underline{\Pi}}$ \cite{LandauLifschitzEngl}.
The temporal change of momentum in an Eulerian control volume is balanced by the net momentum flux through its surface
and by external body forces $\vec{f}_b$ acting on the volume as
\begin{equation}
   \partderivT{ \left( \rho_f \vec{u} \right) } = - \divOp \underline{\underline{\Pi}} + \vec{f}_b.
   \label{Eq:CompMomConservGen}
\end{equation}
The second-order tensor $\underline{\underline{\Pi}}$ comprises a term for the convective transport of momentum
and the total stress tensor $\underline{\underline{\sigma}}$ for the momentum transfer due to pressure and viscosity
\begin{equation}
   \underline{\underline{\Pi}} = \rho_f \, \vec{u}  \vec{u}^\top - \underline{\underline{\sigma}}.
   \label{Eq:MomFluxDensityTensor}
\end{equation}
The total stress tensor $\underline{\underline{\sigma}}$ can be decomposed into a part representing normal stresses related to the pressure
and a viscous part related to shear stresses. 
For incompressible Newtonian fluids, the stress tensor reads as
\begin{equation}
   \underline{\underline{\sigma}} = -p \underline{\underline{I}} + \mu_f \left[ \gradOp \vec{u} + (\gradOp \vec{u})^\top \right],
   \label{Eq:TtlStressTns}
\end{equation}
where the first term with the second-rank identity tensor $\underline{\underline{I}}$ contains the thermodynamic pressure $p$ defined according to Landau \& Lifshitz~\cite{LandauLifschitzEngl} as used in the LBM literature~\cite{dellar2001bulk}.
The second term with dynamic viscosity $\mu_f$ represents the shear stresses that are proportional to the rate of deformation~\cite{happel1983low}. 
With this stress tensor the incompressible Navier-Stokes equation results from \Eqn{Eq:CompMomConservGen}
with the continuity equation for compressible fluids~\cite{Bartuschat:2016:Diss} as
\begin{equation}
   \underbrace{\rho_f \left( \partderivT{ \vec{u}} + (\vec{u} \cdot \gradOp) \vec{u} \right)}_{\substack{\text{inertial forces}}} =
   \underbrace{-\gradOp p}_{\substack{\text{pressure stress}}} + \underbrace{\mu_f \lapOp \vec {u}}_{\substack{\text{viscous stress}}} + 
   \underbrace{\vec{f}_b.}_{\substack{\text{external body force}}}
   \label{Eq:IncompNS_MomentumEq}
\end{equation}
The left-hand side represents inertial forces, and
the right-hand side describes surface forces equivalent to stress in the fluid~\cite{batchelor1979introduction}
and body forces. These volume forces include gravity or electrostatic force
and are represented as force per unit volume $\vec{f}_b$.

An important dimensionless quantity to characterize fluid flows is the Reynolds number $Re = \frac{U L}{\nu_f}$,
with kinematic viscosity $\nu_f = \frac{\mu_f}{\rho_f}$ and characteristic velocity and length scale, $U$ and $L$.
In the creeping motion regime ($Re \ll 1$), inertial forces are negligible, 
and the Stokes equations for incompressible Newtonian fluids result as
\begin{equation}
   \begin{array}{r@{\hspace{0.6ex}}l} 
      -\gradOp p + \mu_f \lapOp \vec {u} + \vec{f}_b &= 0 \\
      \divOp \vec{u} &= 0.
   \end{array}
   \label{Eq:IncompStokes}
\end{equation}
For these linear equations the superposition principle holds, which is often utilized for fluid-particle interaction in Stokes flow
and is employed in \Sect{SubSec:ElectrophorVel_ValidOpenDom}.\\

Particles immersed in a fluid experience a force in case of relative fluid motion or a pressure gradient in the fluid.
This force can be calculated from the stresses in the fluid next to the particle
by integrating the stress tensor $\underline{\underline{\sigma}}$ over the particle surface $\Gamma_p$~\cite{happel1983low,Masliyah:2006:ElectrokinColTransp}
\begin{equation}
   \vec{F}_\text{part} = \int\limits_{\Gamma_p} \underline{\underline{\sigma}} \cdot \vec{n} \; dA.
   \label{Eq:NS_ForceOnPart}
\end{equation}
Here, $dA$ denotes the surface area elements and $\vec{n}$ the associated normal vector.
For simple cases, such as spherical bodies moving in unbounded Stokes flow,
or single bodies moving in confined domains, 
analytical solutions for the particle motion are known as described in the following.
More complex fluid-structure interaction problems must be solved numerically, \eg{} by the LBM.

For the computation of particle motion in incompressible fluids,
hydrostatic effects that result \eg from gravity,
do not have to be considered explicitly in the momentum equation.
In this case, the force exerted by the fluid on the particle according to Eqns.~\eqref{Eq:NS_ForceOnPart} and \eqref{Eq:TtlStressTns}
is the \emph{drag force} given by
$$\vec{F}_d = - \int\limits_{\Gamma_p} p  \; \vec{dA} + \mu_f \int\limits_{\Gamma_p}  \gradOp \vec{u} + (\gradOp \vec{u})^\top \; \vec{dA},$$
where $p$ is the hydrodynamic part of the total pressure.

The resistance to the motion of a sphere in an unbounded fluid at very low Reynolds numbers
can be calculated from the above expression for the drag force
and the Stokes equations \eqref{Eq:IncompStokes}.
For a sphere located at $\vec{x}=\vec{0}$, the unbounded fluid is represented by
$$
      \vec{u} \rightarrow 0 \quad \text{as} \quad \vec{x} \rightarrow \infty
$$
imposed on the fluid velocity in the Stokes equations.
The resulting drag force 
\begin{equation}
   \vec{F}_d = - 6 \pi \mu_f R \vec{U}.
   \label{Eq:StokesDragForce}
\end{equation}
acting on a rigid sphere of radius $R$
and velocity $\vec{U}$ was derived by Stokes~\cite{stokes1851effect}.
This equation is commonly referred to as \emph{Stokes' law}.

For a sphere moving in a fluid subject to a constant force,
Stokes' law relates the terminal steady-state velocity of the sphere to the drag force exerted by the fluid.
Such a force may \eg be the Coulomb force acting on a charged particle.
The terminal sphere velocity is then obtained
from the balance of the external force and the drag force, $\vec{F} + \vec{F}_d = \vec{0}$, as
\begin{equation}
   \vec{U} = \frac{1}{ 6 \pi \mu_f R} \vec{F}\;.
    \label{Eq:StokesVelSphere}
\end{equation}

A particle moving in a confined domain experiences a retardation caused by surrounding walls.
The effect of walls on a moving particle in Stokes regime can be determined by means of the method of reflections,
as described in detail in~\cite{happel1983low}.
Happel \& Bart~\cite{Happel1974SettlingSquareDuct} employed this method to obtain a first-order correction to the drag force on a sphere settling in a long square duct with no-slip walls.
Miyamura et~al.~\cite{miyamura1981experimental} found polynomial expressions for the increased drag
by fitting the coefficients to experimentally obtained settling velocities of spheres in different confining geometries.
The correctness of the wall effect recovered in LBM simulations with the fluid-particle interaction algorithm 
employed in this article was verified in \cite{Bartuschat:2016:Diss} against these expressions.

\section{Electrokinetic Flows \label{Sec:ElkinFlow}}
\subsection{Governing Equations for Electrokinetics}
The transport of ions in fluids subject to electric fields that occurs in electrokinetic flows can be modeled by means of a continuum theory,
similar to fluid dynamics.
Instead of modeling individual ions and their interactions, local ion concentrations $n_i$ and fluxes $\vec{j}_i$ of the different ionic species $i$ are considered.
Based on these macroscopic quantities the ion transport in dilute electrolyte can be described by the 
law for the conservation of ionic species in a solution 
\begin{equation}
   \partderivT{n_i} = - \divOp \vec{j}_{i}
   \label{Eq:IonConserv}
\end{equation}
in the absence of chemical reactions.
Here $n_i$ denotes the number density or concentration that is related to the molar concentration as $c_i = {n_i}/{N_A}$ with Avogadro's number $N_A$.
The total ionic flux $\vec{j}_{i}$ of species $i$ comprises an advective flux with a common mass average velocity $\vec{u}$ for all species
and fluxes relative to the advective flux due to diffusion and electric migration~\cite{Masliyah:2006:ElectrokinColTransp}.
This relation is expressed by the \emph{Nernst-Planck} equation
\begin{equation}
   \vec{j}_{i}   = \underbrace{ n_i \vec{u}                     }_{\substack{\text{advective flux}}}           
                 - \underbrace{ D_i \gradOp n_i                 }_{\substack{\text{diffusive flux}}}           
                 - \underbrace{ n_i \mu^*_i \, \gradOp \Phi,    }_{\substack{\text{migration flux}}}
   \label{Eq:NernstPlanck}
\end{equation}
where $D_i$ and $\mu_i^*$ represent the spatially homogeneous diffusion coefficient and ionic mobility of species $i$, respectively,
and $\gradOp \Phi$ the local electric potential gradient.
The ionic mobility is defined as $\mu_i^* =: \frac{D_i \,  z_i  \, e }{k_B \, T}$, 
where $e$ denotes the elementary charge, $z_i$ the valence of a given ionic species,
$k_B$ the Boltzmann constant and $T$ the temperature.

To model the influence of the charged ions on the electric potential governed by the Poisson equation
\begin{equation}
   - \lapOp \Phi(\vec{x}) = \frac{\rho_e}{\varepsilon_e}
   \label{Eq:ElPoissonUnifPerm}
\end{equation}
for spatially uniform fluid permittivity $\varepsilon_e$,
the ion charge distribution is considered in terms of the local mean macroscopic charge density as 
\begin{equation}
   \rho_e = \sum\limits_i e \, z_i \, n_i.
   \label{Eq:ChrgDensity}
\end{equation}

The Poisson-Nernst-Planck system Eqns.~\eqref{Eq:IonConserv}--\eqref{Eq:ElPoissonUnifPerm} is highly nonlinear, 
and solving the overall system is computationally very expensive, especially for electrophoresis of many particles. 
Therefore the problem is simplified by restriction to equilibrium considerations based on the Boltzmann distribution that capture the dominant electrophoretic effects.
The resulting Poisson-Boltzmann equation holds for \mbox{(quasi-)thermodynamic} equilibrium
when the ion distribution is not affected by fluid flow or by externally applied electric fields.

For the fluid this means that flows must be diffusion-dominated, characterized by very small P{\'{e}}clet numbers $Pe = \frac{U L}{D}$ for mass transfer. 
These dimensionless numbers relate the advection rate of a fluid to its diffusion rate for a given flow speed $U$, length scale $L$  (\ie, particle radius), and diffusion coefficient $D$.
Additionally the applied electric field must be weak compared to the potential difference across the diffuse layer of characteristic thickness or Debye length $\lambda_D$, \ie{}, $E_\text{ext} \ll \frac{\zeta}{\lambda_D}$~\cite{OBrien:White:1978:ElphorMobil}.

Based on these assumptions the electric potential $\psi$ resulting from the non-uniform ion distribution in the EDL is considered in the following, 
instead of the total electric potential $\Phi=\psi+\varphi$ that additionally comprises the potential $\varphi$ of the externally applied electric field.\\
The Boltzmann distribution for ions can be derived from the Nernst-Planck equation, as outlined in~\cite{Masliyah:2006:ElectrokinColTransp}.
Considering the Nernst-Planck equation~\eqref{Eq:NernstPlanck} in one dimension at 
equilibrium and integrating from a reference point in the bulk 
with potential $\psi_\infty$ and concentration $n_{i\infty}$, yields
\begin{equation}   
   n_i = n_{i\infty} \, e^{\displaystyle{-\frac{z_i \, e (\psi - \psi_\infty)}{k_B \, T}}}.
\end{equation}
Setting the reference potential $\psi_\infty$ in the electroneutral bulk solution to zero recovers the Boltzmann distribution
with the number density $n_{i\infty}$ at the location of the neutral state. 

From Poisson's equation~\eqref{Eq:ElPoissonUnifPerm} with net charge density as in \Eqn{Eq:ChrgDensity} 
and the obtained Boltzmann distribution,
the Poisson-Boltzmann equation follows as
\begin{equation}
  - \lapOp \psi = \frac{e}{\varepsilon_e} \sum\limits_i z_i n_{i\infty} e^{\displaystyle{-\frac{z_i \, e \, \psi}{k_B \, T} }},
\end{equation}
relating the electric potential $\psi$ to the ion concentrations at equilibrium.
For binary, symmetric electrolyte solutions comprising two species of valence $z = -z_- = z_+$, the Poisson-Boltzmann equation takes the form
\begin{equation}
- \lapOp \psi = - \frac{2 \, z \, e \, n_{\infty}}{\varepsilon_e} \sinh \left( \frac{z \, e \, \psi}{k_B \, T} \right).
   \label{Eq:SPBE}
\end{equation}

For low $\zeta$-potentials compared to the thermal voltage ${k_B \, T}/{e}$, 
the term $\frac{z \, e \, \psi}{k_B \, T}$ in \Eqn{Eq:SPBE} becomes smaller than unity and the \emph{Debye-H{\"u}ckel approximation} (DHA) can be applied.
At room temperature this is fulfilled for $\zeta \ll \frac{\SI{25.7}{\milli\volt}}{z}$ \cite{Hunter:1981:ZetaPotCollSci}.
In this case the approximation $\sinh(x) \approx x$ is accurate, up to a small error of order $\mathcal O(x^3)$ (\cf{} Taylor's expansion).
With this linearization, the symmetric Poisson-Boltzmann equation simplifies to the DHA
\begin{equation}
   -\lapOp \psi = - \frac{2 \, e^2 \, z^2 \, n_{\infty}}{\varepsilon_r \varepsilon_0 \, k_B \, T} \; \psi = - \kappa^2 \, \psi.
   \label{Eq:DebyHuckApproxPot}
\end{equation}
This equation was originally derived by Debye \& H{\"u}ckel \cite{debye1923ElektrolyteI}
for strong electrolytes \cite{Hunter:1981:ZetaPotCollSci}.
The parameter $\kappa = 1/\lambda_D$, defined by
\begin{equation}
   \kappa := \sqrt{\frac{\varepsilon_r \varepsilon_0 \, k_B \, T}{2 \, e^2 \, z^2 \, n_{\infty}}},
   \label{Eq:DebyHuckParam} 
\end{equation}
is commonly referred to as \emph{Debye-H{\"u}ckel parameter}.
Moreover, the charge density in the fluid is then given by
\begin{equation}
   \rho_e (\psi) = - \kappa^2 \varepsilon_e \psi.
   \label{Eq:ChrgDensDHA}
\end{equation}

\subsection{Analytical Solutions for Electrophoresis}
\begin{figure}[h!]
  \centering
         \includegraphics[scale=1.1]{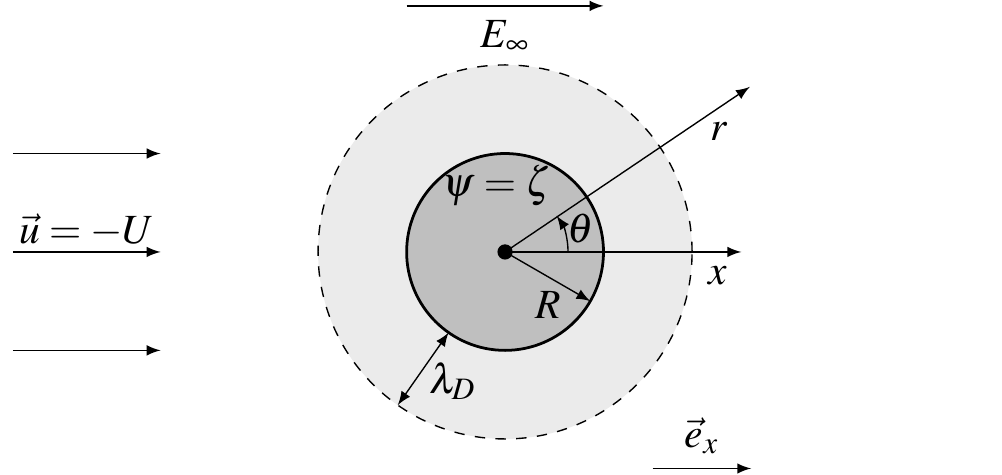} 
    \caption[Electrophoresis setup according to Henry]{Electrophoresis setup of a stationary (negatively) charged sphere of radius $R$,
                                                      surrounded by a double layer and subject to an applied electric field in opposing fluid flow.
                                                      Similar to \cite{probstein1989Physicochem}.
                                                      \label{Fig:Electrophoresis}}
\end{figure}

In this article, we consider spherical particles with uniform $\zeta$-potential distribution as depicted in \Fig{Fig:Electrophoresis}.
The electric potential $\psi$ for such a particle of radius $R$ is represented by the Debye-H{\"u}ckel equation in spherical-polar coordinates as
\begin{equation}
   \frac{1}{r^2} \frac{\textup{d}}{\textup{d}r} \left(r^2 \frac{\textup{d}\psi}{\textup{d}r}  \right) = \kappa^2 \psi,
   \label{Eq:DH_Radial}
\end{equation}
with radial distance $r$ from the sphere center and subject to the Dirichlet BCs
\begin{equation}
\setlength\arraycolsep{6pt}
   \begin{array}{ccc}
   \psi = \zeta &\text{at}& r = R, \\
   \psi \rightarrow 0 &\text{as}&  r \rightarrow \infty.
   \label{Eq:DH_BCS} 
   \end{array}
\end{equation}
Solving this equation subject to these BCs results in the electric potential distribution in the surrounding EDL (and beyond) as \cite{Hunter:1981:ZetaPotCollSci}
\begin{equation}
   \psi (r) = \zeta  \, \frac{R}{r} e^{-\kappa(r-R)} \qquad \text{for } r \geq R.
   \label{Eq:PotDistrSphereDH}
\end{equation}

An analytical solution for the terminal electrophoretic speed $U$ of a single spherical particle at steady state and arbitrary EDL thickness has been derived by Henry~\cite{henry1931cataphoresis}.
Initially the problem was formulated to account for finite conductivities of particle and medium, and the potential in the EDL was described by the Poisson-Boltzmann equation. 
The final results, however, were provided for insulating spheres, and sufficiently low $\zeta$-potentials for the Debye-H{\"u}ckel approximation to hold.
Therefore the Debye-H{\"u}ckel approximation is employed in the following to represent the ion distribution around the particles under electrophoretic motion.

Henry considered a stationary sphere in a steadily moving liquid.
The terminal particle speed was imposed as opposing velocity $-U$
far from the particle to bring the system to rest. As shown in \Fig{Fig:Electrophoresis},
a spherical-polar coordinate system fixed at the particle center with radial distance $r$ and polar angle $\theta$ was used. 
Under the assumption that the electric potential in the EDL is not distorted from its equilibrium distribution by a weak applied field and the fluid flow,
the potentials $\varphi$ and $\psi$ were linearly superimposed.
Therefore, the electric potential $\psi$ in the diffuse double layer is described by the Poisson-Boltzmann equation
and the applied potential $\varphi$ by a Laplace equation.
The BCs for the Laplace equation applied by Henry represent the insulating particle by homogeneous Neumann BCs at the particle surface and 
impose the applied field by the inhomogeneous Neumann condition $\restr{ {\partial \varphi}/{\partial x} }{r \rightarrow \infty} = - E_{\infty}$.
For the Poisson-Boltzmann equation, the $\zeta$-potential at the hydrodynamic radius $R$ and the decaying potential were imposed as given in \Eqn{Eq:DH_BCS}.

Making use of the equations for the electric potential, the Stokes equations 
with body force term on the right-hand side
\begin{eqnarray}
   -\mu_f \lapOp u + \gradOp p &=& -\rho_e \, \gradOp \left( \varphi + \psi \right) \\
   \divOp \vec{u}                      &=& 0,
\end{eqnarray}
were solved by Henry~\cite{henry1931cataphoresis}.
In addition to the BC imposing the opposing velocity 
the no-slip condition $\restr{\vec{u}}{r=R} = 0$ was applied at the particle surface.

From the flow field around the particle, the force acting on the particle was obtained by integrating the normal stresses over the sphere surface.
To the resulting force that comprises Stokes drag and electric components, the electrostatic force on the particle due to its fixed surface charge was added.
The total force must vanish at steady motion and was thus equated with zero, resulting in the electrophoretic velocity
\begin{equation}
   \vec{U}_\text{EP} = \frac{\varepsilon_e}{\mu_f} \underbrace{ \left[ \left( \psi_R + R^3 \left( 5 R^2 \int\limits_\infty^R \frac{ \psi }{r^6} \textup{d}r - 2 \int\limits_\infty^R \frac{\psi }{r^4} \textup{d}r \right)  \right) \right] }_{\substack{ = \zeta \, f\left( \kappa R \right) }} \; \vec{E}_\text{ext}
   \label{Eq:ElPhorVelSphereHenryFull}
\end{equation}
obtained by Henry for an insulating particle subject to an applied field of strength $\vec{E}_\text{ext}$.
The function $f\left( \kappa R \right)$ introduced in~\cite{henry1931cataphoresis}
is usually referred to as Henry's function.
In Ohshima~\cite{ohshima1994simple} the following expression is derived, which approximates the integral equations as
\begin{equation}
   f\left( \kappa R \right) = \frac{2}{3} \left( 1 + \frac{1}{2 \left[ 1+ \dfrac{2.5}{\kappa R \left( 1 + 2 e^{-\kappa R} \right) }  \right]^3 } \right),
    \label{Eq:ElPhorMobilitySphere}
\end{equation}
with a relative error below 1\% for all values of $\kappa R$ \cite{ohshima1994simple}.

With this approximation, Henry's analytical solution for the electrophoretic velocity of a spherical, non-conducting particle
in an unbounded electrolyte solution of dynamic viscosity $\mu_f$ and Debye-H{\"u}ckel parameter $\kappa$ results as
\begin{equation}
   \vec{U}_\text{EP} = \frac{2 \varepsilon_e \zeta}{3 \mu_f} \left( 1 + \frac{1}{2 \left[ 1+ \frac{2.5}{\kappa R \left( 1 + 2 e^{-\kappa R} \right) }   \right]^3 } \right) \; \vec{E}_\text{ext}.
    \label{Eq:ElPhorVelSphere}
\end{equation}
This solution is correct to the first order of the $\zeta$-potential, since the relaxation effect is neglected \cite{Ohshima2006Book}.

For the simulations in this article, the particle charge must be known to compute the electrostatic force on the particle.
Analytical solutions for electrophoretic motion such as Henry's equation, however, are typically given in terms of the $\zeta$-potential,
which is defined at the slip surface between the compact and diffuse EDL layer.
Since the particle charge is acquired as a surface charge, for a given $\zeta$-potential the surface charge enclosed by the slip surface is needed in the simulations.
The surface charge density is hereby obtained from the overall surface charge bound at the fluid-particle interface and in the Stern layer.
This approach is justified by the fact that the electric potential at the Stern surface and the $\zeta$-potential can in general be assumed to be identical~\cite{Shaw1992Colloid}.\\
The relation of the surface density $\sigma_s$ to the $\zeta$-potential is obtained from the
Neumann BC on the surface of the insulating particle \cite{Ohshima2006Book}
\begin{equation}
   \sigma_s = \restr{ -\varepsilon_e \frac{d \psi}{d \vec{r}} }{r=R}
   \label{Eq:SurfChrgVSZetaPot}
\end{equation}
in case these electrical properties do not vary in angular direction.
This condition holds for insignificant permittivity of the insulating particle compared to the fluid permittivity $\varepsilon_e$.
With the spatial distribution of $\psi$ around the spherical particle according to~\Eqn{Eq:PotDistrSphereDH}
the $\zeta$--$\sigma_s$ relationship follows from \Eqn{Eq:SurfChrgVSZetaPot} as
\begin{equation}
   \sigma_s = \frac{q_s}{4 \pi R^2} = \zeta \, \varepsilon_e \left( \frac{1 + \kappa \, R}{R} \right).
   \label{Eq:SurfChrgZetaRelationDebHuck}
\end{equation}

For a spherical particle with an EDL potential $\psi$ described by the spherical symmetric Poisson-Boltzmann equation,
the more general $\zeta$--$\sigma_s$ relationship 
\begin{equation}
\resizebox{0.91\textwidth}{!}{$%
   \sigma_s= \frac{2 \varepsilon_e \kappa k_B T }{z e} \sinh\left( \frac{z e \zeta}{2 k_B T}  \sqrt{ 1 + \frac{1}{\kappa R} \frac{2}{\cosh^2\left(  \dfrac{z e \zeta}{4 k_B T}  \right) }  + \frac{1}{\left(\kappa R \right)^2}  \frac{8 \ln\left( \cosh\left( \dfrac{z e \zeta}{4 k_B T} \right) \right)}{ \sinh^2\left( \dfrac{z e \zeta}{2 k_B T} \right) }  } \right)
   $}
   \label{Eq:SurfChrgZetaRelationOhshima}
\end{equation}
for 1-1 and 2-1 electrolyte solutions is derived in \cite{ohshima1982accurate}.
The relative error of this approximation w.r.t.\ the exact numerical results computed by \cite{Loeb1961EDLSpherPart} is below 1\% for $0.5 \leq \kappa R < \infty $ \cite{Ohshima2006Book}.
Since the latter relationship is derived for the spherical Poisson-Boltzmann equation, it is more general than the relationship~\eqref{Eq:SurfChrgZetaRelationDebHuck} for the Debye-H{\"u}ckel approximation.
Therefore \Eqn{Eq:SurfChrgZetaRelationOhshima} is applied in Secs.~\ref{SubSec:ElectrophorVel_ValidOpenDom} and \ref{SubSec:ElectrophorVelMicrochan}
to compute the particle charge for a given $\zeta$-potential.
\section{Numerical Modeling \label{Sec:NumModeling}}
\subsection{Lattice Boltzmann Method with Forcing \label{SubSec:TRT}}
The LBM is a mesoscopic method for the numerical simulation of fluid dynamics based on kinetic theory of gases.
This method statistically describes the dynamics of ensembles of fluid molecules 
in terms of particle distribution functions (PDFs)
that represent the spatial and velocity distribution of molecules in \textit{phase space} over time.
The temporal and spatial variation of PDFs, balanced by molecular collisions, is described by the Boltzmann equation.
The solution of this equation converges towards the Maxwell-Boltzmann velocity distribution of molecules in local thermodynamic equilibrium.
For small deviations from this equilibrium, the Navier-Stokes equations can be derived from 
the Boltzmann equation by means of a Chapman-Enskog expansion (\cf \cite{hanel2004molekulare,ChapmanCowling90}).

In the LBM, the phase space is discretized into a Cartesian lattice $\Omega_{\textup{dx}} \subset \mathbb{R}^D$ of dimension $D$ with spacing $\textup{dx}$,
and a set of $Q$ discrete velocities $\vec{c}_q \in \mathbb{R}^D, q \in \{1,\ldots,Q\}$.
Moreover, time is discretized as $T_{\textup{dt}} = \{ t_n: n = 0,1,2,\ldots \} \subset \mathbb{R}^+_0$,
with a time increment of $\textup{dt} = t_{n+1} - t_n$.
The velocities $\vec{c}_q$ are chosen such that within a time increment,
molecules can move to adjacent lattice sites or stay at a site.
Associated with each of these velocities is a PDF $f_q: \Omega_{\textup{dx}} \times T_{\textup{dt}} \mapsto \mathbb{R}$.
A forward-difference discretization in time and an upwind discretization in space~\cite{wolf2000lattice}
result in the discrete lattice Boltzmann equation
\begin{equation}
  f_q(\vec{x}_i + \vec{c}_q \textup{dt}, t_n + \textup{dt}) - f_q(\vec{x}_i,t_n) = \textup{dt} \mathcal{C}_q + \textup{dt} F_q,
  \label{Eq:discrLBE_extForce}
\end{equation}
with lattice site positions $\vec{x}_i$, discrete collision operator $\textup{dt} \mathcal{C}_q$, and discrete body-force term $F_q$.
This equation describes the advection of PDFs between neighboring lattice sites
and subsequent collisions.

In general, the collision operator can be represented in terms of the collision matrix $S$ as
$\textup{dt} \mathcal{C}_q = \sum_j S_{q j} \left(f_j-f_j^\text{eq}\right)$ (\cf \cite{dHum1992}),
with the vector $\vec{f} := (f_0, f_1, \ldots , f_{Q-1})^\top$ of the PDFs $f_q$, and $\vec{f}^\text{eq}$ of the equilibrium distributions $f_q^\text{eq}$.
The latter are obtained from a low Mach number expansion of the Maxwell-Boltzmann distribution~\cite{hanel2004molekulare}.
With a representation of the macroscopic fluid density as $\rho_f = \rho_0 + \delta \rho$ in terms of a reference density $\rho_0$ and a density fluctuation $\delta \rho$,
the equilibrium distribution function for the incompressible LBM is derived in \cite{HeLuo:97} as
\begin{equation}
   \hspace{-0.4cm}
   f_q^\text{eq} (\rho_0, \vec{u}) = w_q \left[\rho_f + \rho_0 \left( \frac{(\vec{c}_q \cdot \vec{u})}{c_s^2} + \frac{(\vec{c}_q \cdot \vec{u})^2}{2c_s^4} - \frac{(\vec{u}\cdot \vec{u})}{2c_s^2} \right) \right],
\label{Eq:fequ_Incompr_SRT}
\end{equation}
where `$\cdot$' denotes the standard Euclidean scalar product.
This distribution function depends on $\rho_f$, the macroscopic fluid velocity $\vec{u}$, and lattice-model dependent weights $w_q$.
At each instant of time, $\rho_f$ and $\vec{u}$ are given by moments of PDFs as
\begin{equation}
  \begin{array}{l c}
      \rho_f (\vec{x}_i, t) = \sum\limits_{q} f_q(\vec{x}_i, t), \\
      \vec{u}(\vec{x}_i, t) = \frac{1}{\rho_0} \sum\limits_{q} \vec{c}_{q} f_q(\vec{x}_i, t).
      \label{Eq:MacroscopVel}
  \end{array}
\end{equation}
Moreover, the pressure $p$ is given as $p(\vec{x}_i,t)=c_s^2\rho_f(\vec{x}_i,t)$ according to the equation of state for an ideal gas.
We employ the D3Q19 model of \cite{1992:QianBGK} with thermodynamic speed of sound $c_{s}= c/\sqrt{3}$ for the lattice velocity ${c = \textup{dx} / \textup{dt}}$.
For this model, the weights $w_q$ are: ${w_1 = 1/3}$, ${w_{2, \ldots, 7} = 1/18}$, and $w_{8,\ldots, 19} = 1/36$.
As discussed in \cite{HeLuo:97},
$f_q^\text{eq}$ recovers the incompressible Navier-Stokes equation
with at least second-order accuracy of the Mach number ${\mbox{Ma} := \frac{|\vec{u}|}{c_s}}$, $\mathcal O(\mbox{Ma}^2)$.
In LBM simulations, the kinematic fluid viscosity $\nu_f$ is generally determined by a dimensionless relaxation time $\tau$ of the collision operator as
\begin{equation}
   \nu = \left(\tau - \frac{1}{2}\right) c_{s}^2 \textup{dt}.
   \label{Eq:kinVisc}
\end{equation}
As shown in \cite{Sterling1996_StabilLBM}, with this definition the LBM is second-order accurate in space and time.\\
Among the different collision operators available for the LBM, we employ the
two-relaxation-time (TRT) collision operator of \cite{ginzburg2004lattice,ginzburg2008two}
\begin{equation}
\sum\limits_{j} S_{q j} \left( f_j - f_j^{\text{eq}} \right) =
    \lambda_e \left( f^e_q - f_q^{\text{eq},e} \right) +
    \lambda_o \left( f^o_q - f_q^{\text{eq},o} \right),
  \label{Eq:TRTOp}
\end{equation}
with the relaxation parameters $\lambda_e$ and $\lambda_o$ for even- and odd-order non-conserved moments, respectively.
Alternative collision operators either have disadvantages regarding stability and accuracy \cite{Luo2011CollModEff} such as the BGK model \cite{1954:BGKPhysRev.94.511}
or are computationally more costly such as the MRT operator \cite{dHum1992} or the cumulant operator \cite{Geier2015Cumulant}.
To ensure stability, both relaxation parameters should be within the interval ${]-2,0[}$, \cf~\cite{higuera1989lattice,ginzburg2008two}.
The even relaxation parameter is related to the dimensionless relaxation time by ${\lambda_e=-\frac{1}{\tau}}$
and therefore determines the fluid viscosity.
The free parameter $\lambda_o$ is set to $\lambda_o = - 8  (2-1/\tau)/(8-1/\tau)$ in this article,
which prevents $\tau$-dependent boundary locations for no-slip BCs as they arise for the BGK operator.
Instead, walls aligned with the lattice dimensions are fixed half-way between two lattice sites, as shown in \cite{ginzburg2008study}.
For the TRT operator, the PDFs are decomposed as $f_q = f^e_q + f^o_q$ into their even and odd components
\begin{equation}
  \begin{array}{l c r}
  f^e_q = \frac{1}{2} ( f_q + f_{\bar{q}} ) &\text{ and }& f^{\text{eq},e}_q = \frac{1}{2} ( f^\text{eq}_q + f^\text{eq}_{\bar{q}} ) \\
  f^o_q = \frac{1}{2} ( f_q - f_{\bar{q}} ) &\text{ and }& f^{\text{eq},o}_q = \frac{1}{2} ( f^\text{eq}_q - f^\text{eq}_{\bar{q}} ),
  \end{array}
\end{equation}
with $\vec{c}_q = - \vec{c}_{\bar{q}}$.
The local equilibrium distribution function in \Eqn{Eq:fequ_Incompr_SRT} is then given by
\begin{equation}
  \begin{array}{l c}      
      f^{\text{eq},e}_q = w_q \left( \rho_f - \frac{\rho_0}{2c_{s}^2}({\vec{u}} \cdot \vec{u}) + \frac{\rho_0}{2c_{s}^4}({\vec{c}_q} \cdot \vec{u})^2 \right) \\
      f^{\text{eq},o}_q = w_q \frac{\rho_0}{c_{s}^2}(\vec{c}_q \cdot \vec{u}).
  \end{array}
  \label{Eq:EqPDF_TRT}
\end{equation}

At each time step $t_n \in T_{\textup{dt}}$ the lattice Boltzmann method performs a \emph{collide--} and a \emph{stream step}
\begin{equation}
  \hspace{-0.2cm}
  \begin{array}{r@{\hspace{0.5ex}}l}
  \tilde{f}_q(\vec{x}_i,t_n) = f_q(\vec{x}_i,t_n) &+ \lambda_e [ f^e_q(\vec{x}_i,t_n) - f_q^{\text{eq},e}(\vec{x}_i,t_n) ] \\ 
                                                  &+ \lambda_o [ f^o_q(\vec{x}_i,t_n) - f_q^{\text{eq},o}(\vec{x}_i,t_n) ]
  \label{Eq:LBMcollideStep}                                                 
 \end{array}
\end{equation}
\begin{equation}
f_q(\vec{x}_i + \vec{e}_q, t_n+\textup{dt}) = \tilde{f}_q(\vec{x}_i,t_n) + \textup{dt} F_q,
  \label{Eq:LBMstreamStep}                                                 
\end{equation}
where $\tilde{f}_q$ denotes the post-collision state and $\vec{e}_q =\vec{c}_q \textup{dt}$ a discrete lattice direction.

In the stream step, the product of $\textup{dt}$ and the forcing term $F_q$ is added to the post-collision PDFs.
The term $F_q$ considers the external effect of body forces acting on the fluid.
In this article, the discrete forcing term according to Luo~\cite{1997_Luo_AnalytSolnLinearizedLBE} is employed as
\begin{equation}
  F_{q} = w_{q} \left[\frac{(\vec{c}_{q}-\vec{u})}{c_{s}^2} + \frac{(\vec{c}_{q}\cdot\vec{u})}{c_{s}^4}\vec{c}_{q}\right] \cdot \vec{f}_b,
  \label{Eq:LBE_forcetermLuo}
\end{equation}
with $\vec{f}_b$ representing the electrical body force per unit volume.
The forcing terms lead to an additional term in the continuity equation that arises for spatially varying external forces,
as shown in~\cite{Ladd:2001:LBsimSuspLubr,GuoForcing2002}.
This additional term can be removed by incorporating the external force in the momentum density definition as
\begin{equation}
   \vec{u} = \frac{1}{\rho_0} \left( \sum_q f_q \vec{c}_{q} + \frac{\textup{dt}}{2} \vec{f}_b \right).
 \label{Eq:modified_velocity}
\end{equation}
Thus, \cite{Ladd:2001:LBsimSuspLubr} suggest to use the forcing term~\eqref{Eq:LBE_forcetermLuo} 
in combination with the modified momentum density definition in~\Eqn{Eq:modified_velocity} for the BGK.
We therefore use this forcing term with second-order accuracy,
together with the modified momentum density for the resulting velocity $\vec{u}$.

To increase the computational efficiency of the implementation, the compute-intensive collide step and the memory-intensive stream step are fused to a \emph{stream-collide} step.
In the simulations presented in this article, no-slip and free-slip BCs are applied
as described in \cite{Bartuschat:2014:Tumbling}.

\subsection{Momentum Exchange Approach \label{SubSec:MEA}}
To model the fluid-particle interaction and the resulting hydrodynamic interactions of the particles,
the momentum exchange approach suggested by \cite{nguyen2002lubrication} is employed.
The implementation of this approach in \walberla has recently been applied to simulate fluid-particle interactions also at Reynolds numbers beyond the Stokes regime, as presented in \cite{Rettinger2017comparative,Fattahi2016LaminarTurbuLBM,bogner2014drag}.

For the momentum exchange approach, the solid particles are mapped onto the lattice by considering each cell whose center is overlapped by a particle a moving obstacle cell.
All other lattice cells are fluid cells, resulting in a staircase approximation of the particle surfaces.
These surfaces are represented by surface cells indicated by subscript $s$ in the following. On the fluid cells denoted by subscript $F$, the LBM is applied.
To model the momentum transfer from the particles to the fluid, the velocity bounce-back BC
\begin{equation}
   f_{\bar{q}} \left( \vec{x}_{F}, t_n + \textup{dt} \right) = \tilde{f}_q \left( \vec{x}_{F} , t_n \right) - 2 \frac{\omega_q}{c_s^2} \rho_{0} \vec{c}_{q} \cdot \vec{u}_{s}
   \label{Eq:LBMMovWall}
\end{equation}
is applied at fluid cells with position $\vec{x}_{F} = \vec{x}_s + \vec{e}_{\bar{q}}$ adjacent to a surface cell at $\vec{x}_s$.
This boundary condition introduced in \cite{Ladd_1993_PartSuspPt1} matches the fluid velocity to the local particle surface velocity $\vec{u}_{s}$.

From the sum of all force contributions due to 
the momentum transfer from fluid cells to neighboring surface cells (\cf \cite{Bartuschat:2014:Tumbling}),
the overall hydrodynamic force on the particle can be obtained according to \cite{nguyen2002lubrication} as
\begin{equation}
   \hspace{-0.2cm}
   \vec{F}_{h} = \sum\limits_{s} \sum\limits_{q \in D_s } \left[ 2 \tilde{f}_q \left( \vec{x}_{F}, t_n \right) - 2 \frac{\omega_q}{c_s^2} \rho_{0} \vec{c}_{q} \cdot \vec{u}_{s} \right] \vec{c}_q  \frac{\textup{dx}^3}{\textup{dt}}.
   \label{Eq:LBMhydrodynForce}
\end{equation}
Here, $D_s$ is the set of direction indices $q$ in which a given particle surface cell $s$ is accessed from an adjacent fluid cell.
Analogously, the overall torque $\vec{M}_{h}$ is given by substituting $\vec{c}_q \times \left( \vec{x}_{s} - \vec{x}_{C} \right)$
for the last $\vec{c}_q$ term in \Eqn{Eq:LBMhydrodynForce}, with $\vec{x}_{C}$ representing the particle's center of mass.

The mapping of the solid particles onto the lattice results in fluid cells appearing and disappearing as the particles move.
Therefore, at uncovered lattice sites the PDFs must be reconstructed.
We set the PDFs at those fluid cells to the equilibrium distribution $f^\text{eq} \left( \rho_{0}, \vec{u}_{s} ( \vec{x}_s(t_{n} - \textup{dt}) \right)$ according to~\Eqn{Eq:fequ_Incompr_SRT}
dependent on the local particle surface velocity from the previous time step.

\subsection{Finite Volume Discretization for Electric Potential Equations \label{SubSec:FV_PBE}}
To solve the Debye-H{\"u}ckel approximation
a cell-centered finite volume scheme is applied on the Cartesian lattice $\Omega_\textup{dx}$ introduced in \Sect{SubSec:TRT}.
Associated with this lattice of spacing $\textup{dx}$ that represents the computational domain $\Omega \subset \mathbb{R}^3$
is the three-dimensional cell-centered grid $G_\textup{dx}$ defined (\cf\ \cite{wesseling2004introduction}) as
\begin{equation}
   G_\textup{dx} := \left\{ \ \vec{x}_i \in \Omega \ \Big| \ \vec{x}_i = \begin{pmatrix}j-1/2\\k-1/2\\m-1/2\end{pmatrix}\textup{dx} \ \Big| \ i \in {\textstyle\bigwedge_i}, \ (j,k,m) \in {\textstyle\bigwedge_J} \ \right\},
   \label{Eqn:FineLattice}
\end{equation}
with $\Omega_\textup{dx} = \Omega \, \cap \, G_\textup{dx}$.
For indexing of lattice cells by tuples $(j,k,m)$ of cell indices in the three spatial dimensions, 
the index set ${\textstyle\bigwedge_J} := \{ (j,k,m) \mid \ j=1,\dots,l_\text{x};\; k=1,\dots,l_\text{y};\; m=1,\dots,l_\text{z} \}$
is introduced.
Here $l_\text{x}$, $l_\text{y}$, and $l_\text{z}$ represent the numbers of cells in x, y, and z-direction, respectively.
The set ${\textstyle\bigwedge_J}$ is related to the set of single cell indices 
${\textstyle\bigwedge_i} := \left\{ i \ \mid \ i=1,\dots,l_\text{x} \cdot l_\text{y} \cdot l_\text{z} \right\}$
used for the LBM by a bijective mapping $g: {\textstyle\bigwedge_i} \rightarrow {\textstyle\bigwedge_J}$,
according to \Eqn{Eqn:FineLattice}.

The finite volume discretization of the Debye-H{\"u}ckel approximation \Eqn{Eq:DebyHuckApproxPot}
includes volume integration over each lattice cell
$\Omega_{i} = \Omega_{klm}:=\left[\vec{x}_{j-1,k,m}, \, \vec{x}_{j,k,m} \right] \times \linebreak[1] \left[\vec{x}_{j,k-1,m}, \, \vec{x}_{j,k,m} \right] \times \left[\vec{x}_{j,k,m-1}, \, \vec{x}_{j,k,m} \right]$
and applying the divergence theorem to the
Laplace operator ${\lapOp = \divOp \gradOp}$, 
resulting in
\begin{equation}
   - \oint\limits_{\partial\Omega_{i}} \, \gradOp \psi(\vec{x}) \; \textup{d}\vec{\Gamma}_{i} + \kappa^2 \, \int\limits_{\Omega_{i}} \psi(\vec{x}) \; \textup{d}\vec{x} = 0  \quad \forall \; \Omega_{i} \in \Omega_\textup{dx}. 
   \label{Eq:DiscrFV}
\end{equation}
Here, $\partial\Omega_{i}$ denotes the closed surface of the cell, and $\vec{\Gamma}_{i}$ is a surface directed element.
The cell surface consists of six planar faces with constant outward unit normal vectors $\vec{n}_{{i}_q}$ (${q=1,\ldots,6}$). 
Therefore, the surface integral can be decomposed into a sum of integrals \cite{knabner2003numerical} as
\begin{equation}
   - \oint\limits_{\partial\Omega_{i}} \, \gradOp \psi(\vec{x}) \; \textup{d}\vec{\Gamma}_{i}  = - \sum\limits_{q=1}^6 \; \int\limits_{\partial\Omega_{{i}_q}} \, \gradOp \psi(\vec{x}) \cdot \vec{n}_{{i}_q} \; \textup{d}\Gamma_{{i}_q},
   \label{Eq:DiscrFVInt}
\end{equation}
where $q$ represents LBM direction indices introduced in \Sect{SubSec:TRT},
and $\partial\Omega_{{i}_q}$ 
is the common face with the neighboring cell in direction $q$.
The gradients $\gradOp \psi(\vec{x}_i) \cdot \vec{n}_{{i}_q}$ in normal direction of the faces $\partial\Omega_{{i}_q}$
are approximated at the face centers by central differences of $\psi$ from a neighboring and the current cell as
\begin{equation}
  \restr{\gradOp \psi(\vec{x}_i) \cdot \vec{n}_{{i}_q}}{\vec{x}_i + \frac{1}{2}\,\vec{e}_{q}} \approx \frac{\psi(\vec{x}_i +  \vec{e}_{q} ) - \psi(\vec{x}_i) }{\textup{dx}}.
   \label{Eq:NormalFlux}
\end{equation}
Here, $\vec{e}_{q}$ represents the corresponding lattice direction introduced in \Sect{SubSec:TRT}.\\
Substituting the approximation of the normal fluxes across the surfaces of area $\textup{d}\Gamma_{{i}_q} = \textup{dx}^2,$ \Eqn{Eq:NormalFlux}
into \Eqn{Eq:DiscrFVInt} results in
\begin{equation}
   - \oint\limits_{\partial\Omega_{i}} \, \gradOp \psi(\vec{x}) \; \textup{d}\vec{\Gamma}_{i}  \approx - \sum\limits_{q=1}^6 \; \frac{\psi(\vec{x}_i +  \vec{e}_{q} ) - \psi(\vec{x}_i) }{\textup{dx}} \; \textup{dx}^2.
   \label{Eq:DiscrFVPoisson}
\end{equation}

Applying the above finite volume discretization to the linear term of the Debye-H{\"u}ckel approximation results in 
\begin{equation}
   \kappa^2 \, \int\limits_{\Omega_{i}} \psi(\vec{x}) \; \textup{d}\vec{x} \; \approx \; \kappa^2 \; \psi(\vec{x}_i) \; \textup{dx}^3.
   \label{Eq:DiscrFV_DHTerm}
\end{equation}
This additional term enters the central element of the resulting seven-point stencil $\varXi^\text{DHA}_{\textup{dx}}$ as
\begin{equation}
  \frac{1}{{\textup{dx}}^2}
  \left[
   \begin{matrix}
      \begin{pmatrix} 
        0 & ~0 & ~0 \\
        0 & -1 & ~0 \\
        0 & ~0 & ~0      
      \end{pmatrix}
      \begin{pmatrix} 
        ~0 & -1 & ~0 \\
        -1 & ~6+\kappa^2 \,{\textup{dx}}^2 & -1 \\
        ~0 & -1 & ~0
      \end{pmatrix}
      \begin{pmatrix} 
        0 & ~0 & ~0 \\
        0 & -1 & ~0 \\
        0 & ~0 & ~0      
      \end{pmatrix}
   \end{matrix}
  \right],
\end{equation}
and the right-hand side for each unknown is zero.

\subsection{Parametrization for Electrokinetic Flows \label{SubSec:Parametriz}}
Numerical simulations are typically performed in terms of dimensionless parameters.
To ensure consistently good numerical accuracy independent of the simulated scales,
the physical quantities are mapped to a computationally reasonable numerical value range.
In LBM simulations usually the quantities are expressed in lattice units.
Therefore, physical values must be converted to lattice values before the simulation
and vice versa to obtain physical values from simulation results.
In the following the lattice unit system employed in this article is presented,
providing a common parameterization also for further (electrokinetic) flow scenarios \cite{Bartuschat:2016:Diss}.

Physical simulation parameters are given in terms of the \emph{international system of quantities} (ISQ)
associated to the \emph{international system of units} (SI \cite{SI2006Bureau}) \cite{ISO80000Pt1}.
The SI system comprises the \emph{base units} displayed in \Tab{Tab:SI_LatticeUnits}, together with the corresponding mutually independent 
\emph{base quantities} length, time, mass, electric current, thermodynamic temperature, amount of substance, and luminous intensity.
Moreover, \emph{derived units} such as $\si{\newton} = \si{\kilo\gram \meter \per \square \second}$ are defined
as products of powers of base units \cite{SI2006Bureau}. 

For LBM simulations the base quantities are length, time, and mass density.
\begin{table}[h]
   \begin{center}
      \caption[]{Physical and LBM base quantities for electrokinetic simulations. \label{Tab:SI_LatticeUnits}}
      \begin{tabular}{l|c@{\hspace{1.5ex}}c@{\hspace{1.5ex}}c@{\hspace{1.5ex}}c@{\hspace{1.5ex}}c@{\hspace{1.5ex}}c@{\hspace{1.5ex}}c}
\toprule
         kind of quantity  & length                & time               & mass            & electr.                           & thermodyn.      & chem.        & photometr.      \\
\midrule
         ISQ quantity      & $\textup{x}$          & $\textup{t}$       & $m$             & $I$                               & $T$             & $n$          & $I_v$           \\
         SI unit           & \si{\meter}           & \si{\second}       & \si{\kilo\gram} & \si{\ampere}                      & \si{\kelvin}    & \si{\mole}   & \si{\candela}   \\
\midrule
         LBM quantity      & $\textup{x}$          & $\textup{t}$       & $\rho$          & $\Phi$                            & --              & --           & --              \\
         lattice unit      & $\textup{dx}$         & $\textup{dt}$      & $\rho_{0}$      & $\si{\volt}$                      & --              & --           & --              \\
\bottomrule
      \end{tabular}
   \end{center}
\end{table}
The corresponding \emph{lattice base units} are the (physical) lattice spacing $\textup{dx}$, time increment $\textup{dt}$, and fluid reference density $\rho_{0}$.
Therefore, the numerical values of these quantities in \emph{lattice units} (LUs) become unity, as shown in \Tab{Tab:UnitConversion}.
The lattice parameters representing these dimensionless numerical values in LUs are indicated with subscript $L$ in the following, \eg $\textup{dx}_L$ for the lattice spacing.
Performing LBM computations on such normalized lattice parameters saves numerical operations:
Additional scaling factors are avoided, \eg in the LBM stream-collide step,
when computing the equilibrium distribution function given in \Eqn{Eq:fequ_Incompr_SRT} (\cf $c_L = \textup{dx}_L/\textup{dt}_L =1$)
or the macroscopic velocity according to \Eqn{Eq:MacroscopVel} (\cf $\rho_{0,L}=1$).

For the electrokinetic simulations, the electric potential $\Phi$
is chosen as electric base quantity.
This quantity, however, is not scaled on the lattice but keeps its numerical value that typically lies in a range not too far from unity.
In contrast to the SI system, the LU system for electrokinetic simulations requires no base units
corresponding to the thermodynamic temperature $T$ or the amount of substance $n$:\\
In the simulations, temperature appears only in combination with the Boltzmann constant as energy, \ie{}, $\mathpzc{E} = k_B \, T$,
with the derived unit $\left[ \mathpzc{E} \right] = [m] \, [\textup{x}]^2 / [\textup{t}]^2$.
With the relation of mass and mass density ${\left[ m \right] = \left[ \rho_0 \right] \, [\textup{x}]^3}$, this unit
can be represented in lattice base units (see \Tab{Tab:SI_LatticeUnits}).\\
Moreover, by representing the molar concentration with unit $\left[ c_i \right] = \si{\mole\per\litre}$ 
in terms of the number density as $n_i=c_i \, N_A$ with Avogadro's number $N_A = \SI{6.02214e23}{\per\mole}$,
the unit `$\si{\mole}$' cancels out, yielding $[n_i] = [\textup{x}]^{-3}$.

With the choice of the potential $\Phi$ as base quantity, the electric current $I$ becomes a derived quantity in the LU system.
The unit of $I$ can be derived from 
the energy in electrical units $\left[ \mathpzc{E} \right] = [I] \, [\Phi] \, [\textup{t}]$ and
the above definition of energy in terms of lattice units $\left[ \mathpzc{E} \right] = \left[ \rho_0 \right] \, [\textup{x}]^5 / [\textup{t}]^2$.
Equating both relations results in the derived lattice unit 
\begin{equation}
   [I] = \frac{ [\rho_0] \, [\textup{x}]^5 }{ [\Phi] \, [\textup{t}]^3 }.
   \label{Eq:DerivedUnitAmpere}
\end{equation}

In \Tab{Tab:UnitConversion} different physical quantities and their SI units are displayed,
as well as their representation by (dimensionless) lattice parameters and their lattice units.
\begin{table}[h!]
   \begin{center}
      \caption[]{ Relation of physical quantities and lattice parameters for electrokinetic simulations.
                  
                  \label{Tab:UnitConversion}} 
      \begin{tabular}{l|c|c@{\hspace{1ex}}r@{\hspace{0.2ex}}l|c}
\toprule
      physical quantity                   & SI unit                                 & lattice parameter  & \multicolumn{2}{c|}{numerical value}                            & lattice unit                                       \\
\midrule
      $\textup{dx}$  (lattice spacing)    & \si{\meter}                             & $\textup{dx}_L$    & $\frac{1}{\textup{dx}}$                                         & $\,\textup{dx}$ \footnotesize{(= 1)}  & ${\textup{dx}}$                       \\[0.5ex] 
      $\textup{dt}$  (time increment)     & \si{\second}                            & $\textup{dt}_L$    & $\frac{1}{\textup{dt}}$                                         & $\,\textup{dt}$ \footnotesize{(= 1)}  & ${\textup{dt}}$                       \\[0.5ex]
      $\rho_0$       (fluid density)      & \si{\kilo\gram\per\cubic\meter}         & $\rho_{0,L}$       & $\frac{1}{\rho_0} $                                             & $\,\rho_0$ \footnotesize{(= 1)}       & ${\rho_{0}}$                          \\[0.5ex]
      $\Phi$         (electr. potential)  & \si{\volt}                              & $\Phi_L$           & $\frac{1}{\si{\volt}}$                                          & $\,\Phi$                       & ${\si{\volt}}$                        \\[0.5ex]
\hline
\rule{0ex}{2.4ex} \hspace{-2mm}  
      $L$            (length)             & \si{\meter}                             & $L_L$              & $\frac{1}{\textup{dx}}$                                         & $\,L$                      & ${\textup{dx}}$                       \\
      $\nu$          (kinem. viscosity)   & \si{\square\meter\per\second}           & $\nu_L$            & $\frac{\textup{dt}}{\textup{dx}^2}$                             & $\,\nu$                    & $\frac{\textup{dx}^2}{\textup{dt}}$   \\[0.5ex]
      $\vec{u}$      (velocity)           & \si{\meter\per\second}                  & $\vec{u}_L$        & $\frac{\textup{dt}}{\textup{dx}}$                               & $\,\vec{u}$                & $\frac{\textup{dx}}{\textup{dt}}$                      \\[0.5ex]
      $m$             (mass)              & \si{\kilo\gram}                         & $m_L$              & $\frac{1}{\rho_{0} \, \textup{dx}^3}$                           & $\,m$                      & ${\rho_{0} \, \textup{dx}^3}$      \\[0.5ex]
      $\vec{F}$       (force)             & \si{\kilo\gram\meter\per\square\second} & $\vec{F}_L$        & $\frac{\textup{dt}^2}{\rho_{0} \, \textup{dx}^4}$               & $\,\vec{F}$                & $\frac{\rho_{0} \, \textup{dx}^4}{\textup{dt}^2}$      \\
      $I$             (electr. current)   & \si{\ampere}                            & $I_L$              & $\frac{\si{\volt} \, \textup{dt}^3}{\rho_{0} \, \textup{dx}^5 }$& $\,I$                      & $\frac{\rho_{0} \, \textup{dx}^5 }{\textup{dt}^3 \, \si{\volt} }$  \\
      $e$             (elem. charge)      & \si{\ampere\second}                     & $e_L$              & $\frac{\si{\volt} \, \textup{dt}^2}{\rho_{0} \, \textup{dx}^5 }$& $\,e$                      & $\frac{\rho_{0} \, \textup{dx}^5 }{\textup{dt}^2 \, \si{\volt} } $  \\
      $\varepsilon_0$ (vac. permittivity) & \si{\ampere\second\per\volt\per\meter} & $\varepsilon_{0,L}$ & $\frac{\si{\volt}\,\textup{dt}^2}{\rho_{0} \, \textup{dx}^5 }$ & $\varepsilon_0$ & $\frac{\rho_{0} \, \textup{dx}^5 }{\si{\volt}\,\textup{dt}^2}$ \\[0.5ex]
      $\vec{E}$ (electr. field )          & \si{\volt\per\meter}                            & $\vec{E}_L$        & $\frac{\textup{dx}}{\si{\volt}}$                 & $\,\vec{E}$                & $\frac{ \si{\volt} }{\textup{dx}}$                     \\[0.5ex]
      $\mathpzc{E}$   (energy)            & \si{\joule}                             & $\mathpzc{E}_L$    & $\frac{\textup{dt}^2}{\rho_0 \, \textup{dx}^5}$                    & $\,\mathpzc{E}$ & $\frac{ \rho_0 \, \textup{dx}^5 }{\textup{dt}^2}$                 \\
\bottomrule
      \end{tabular}
   \end{center}
\end{table}
The conversion of physical quantities to lattice units requires their division by powers of the LBM base quantities with the corresponding SI units.
Since the potential $\Phi$ has the same numerical value in physical and lattice units, the LBM base unit of the potential is simply `\SI{1}{\volt}'.
Therefore, the derived lattice unit `Ampere' for the electric current is given by $\si{\ampere} = \frac{\rho_0 \, \textup{dx}^5}{ \si{\volt} \, \textup{dt}^3 }$ (see \Eqn{Eq:DerivedUnitAmpere}).
The corresponding scale factors for converting the physical parameters (\eg $\nu$)
to the associated lattice parameters (\eg $\nu_L$) are shown in \Tab{Tab:UnitConversion}.
Multiplication with the inverse scale factors converts the lattice parameters back to physical quantities. 
\section{Extension of \walberla for Electrophoresis Simulations \label{Sec:ExtWalberlaElph}}
Electrophoresis simulations require the mutual coupling of fluid dynamics, rigid body dynamics, and electro-statics, as shown in \Fig{fig:EKF_Multiphysics_Interactions}.
In addition to electrostatic and hydrodynamic interactions, the applied field acts on the EDL charge and thereby affects fluid flow and particle motion.
For the electrophoresis algorithm presented below, the equilibrium description of the EDL potential in terms of the linear Debye-H{\"u}ckel equation is employed.
Therefore, the predominant retardation effect is recovered in the simulations.
The applied potential $\varphi$ and the EDL potential $\psi$ are linearly superimposed (\cf Henry's equation, \Sect{Sec:ElkinFlow}),
which is valid for weak applied fields when the EDL distortion by the field is negligible.
Thus, the applied electric field can be imposed directly, without solving the associated Laplace equation.
In the following the main concepts of the \walberla framework are described, together with the functionality for electrophoresis simulations implemented therein.

\subsection{Design Concepts of \walberla \label{Sec:waLBerlaConcept}}
\Walberla is a framework for massively parallel multiphysics simulations with a MPI-based distributed memory parallelization
that is specifically designed for supercomputers.
The main software design goals of this framework are flexibility to combine models of different effects, 
extensibility to allow the incorporation of further effects and details, \eg, for electrokinetic flows, 
and generality to support further applications~\cite{Bartuschat:2016:Diss}.
These goals are reached by integrating the coupled simulation models into \walberla
in a modular fashion that avoids unnecessary dependencies between the modules.
This way, the modules can be augmented by more sophisticated models or models tailored to a certain application,
and functionality from different modules can be combined flexibly.
The modular code structure also provides excellent maintainability, since modifications of the code in one module do not affect other modules.
Developers can therefore efficiently locate faulty modules and find bugs inside these modules, also by systematically utilizing automatic tests.

In addition to \emph{modules}, the \walberla code structure comprises a \emph{core} for sequence control that 
initializes data structures, performs the time stepping, and finalizes the simulation on each parallel process.
By means of \emph{applications}, multiphysics simulations can be defined by assembling the associated functionality and coupled models from the modules.
The coupling strategy for multiphysics simulations is based on accessing mutually dependent data structures (see \cite{Bartuschat:2014:CP} for more details). 
These data strucutres are defined in the modules that implement models for the different physical effects.
Also infrastructural and utility functionality is encapsulated in modules, \eg, for domain setup, MPI communication, BC handling, parameterization, or simulation data output.
For parallel simulations the discretized simulation domain is decomposed into equally sized \emph{blocks} of cells that can be assigned to different parallel processes.
In this \emph{block concept}, each block contains a layer of surrounding ghost cells that is needed for BC treatment and parallelization.
For parallelization, cell data of neighboring processes is copied to the ghost layer, dependent on the data dependencies of the unknowns located on a given block.
Moreover, metadata of a block specifies its location in the simulation domain or its rank for MPI communication.
The \emph{communication concept} provides a simple and flexible communication mechanism tailored to simulations on Cartesian grids and facilitates various communication patterns.
Individual work steps of a simulation algorithm are specified as \emph{sweeps} that are executed on a block-parallel level.
The \emph{sweep concept} defines a structure in which callable objects (\ie kernels) implemented in the modules can be specified at compile time.
By means of dynamic application switches, specific kernels tailored to a given computer architecture or implementing a desired model variant, can be selected at run time.
For time-dependent simulations, the sweeps are organized in a \emph{timeloop} that specifies the order in which the individual work steps are executed at each time step.
To facilitate iterative solvers, sweeps can also be nested to repeatedly perform a grid traversal until a termination criterion is met.

The \emph{boundary condition concept} for handling multiple physical fields, numerical methods, and governing equations, introduced in \cite{Bartuschat:2014:CP},
is applied in this article for moving obstacles with electric BCs.
This concept relies on \emph{flags} to indicate for each boundary cell the kind of boundary treatment that to is be performed, with an individual \emph{BC} flag for each condition.
Moreover, cells adjacent to a boundary are indicated with a \emph{nearBC} flag and non-boundary cells with a \emph{nonBC} flag.
Individual sets of these flags are defined for each governing equation.
Due to specific LBM requirements, the boundary handling is performed such that the BCs are fulfilled when a boundary cell is accessed from a \emph{nearBC} cell in the susequent sweep.
The abstract boundary handling is implemented in the \emph{bc} module and provides functionality for all BCs
are handled either as direct or direction-dependent treatment.
In direction-dependent treatment the BC value is set at a boundary cell dependent on the value at a neighboring cell, 
whereas in direct BC treatment the BC value is directly set at a boundary cell \cite{Bartuschat:2014:CP}.
The actual boundary handling functionality is implemented in corresponding BC classes whose functions are executed when the associated \emph{BC} flag is found.

The \emph{parameterization concept} for multiphysics simulations introduced in \cite{Bartuschat:2016:Diss}
is based on the conversion of physical parameters to lattice units before the simulation, as described in \Sect{SubSec:Parametriz}.
This approach ensures consistent parameters in all \walberla modules, independent of the underlying physics.
Individual modules can therefore be developed independently w.r.t.\ the common lattice unit system.

Simulation parameters and BCs are typically provided to \walberla through an input file.
To ensure a consistent parameter set and a correct mapping of the physical quantities to lattice units,
the class \emph{PhysicalCheck} has been introduced in \walberla\ in \cite{Donath:2011:Diss}.
This class checks the simulation parameter set specified in the input file for completeness and physical validity and converts the parameters to lattice units.
Since the quantities are converted based on the SI system,
the unit Ampere is re-defined for PhysicalCheck according to \Eqn{Eq:DerivedUnitAmpere} to support simulations including electric effects.

\subsection{Overview of \walberla Modules for Electrophoresis Simulations}
In the following an overview of the modules relevant for electrophoresis simulations is given.
For fluid simulations with the LBM, the \emph{lbm} module implements various kernels 
for the stream-collide step with the different collision operators and forcing terms described in \Sect{SubSec:TRT}.
The classes for treating the corresponding BCs are provided by the associated \emph{lbm\_bc} module.
In the \emph{lbm} module block-fields of cells are provided for the PDFs, the velocity, the density, and an external force.
The external force field is used for coupling the LBM to other methods,
\eg via the forces exerted by electric fields on the EDL and the fluid in electrophoresis simulations.
The PDF and velocity field are accessed by \emph{moving\_obstacle} module functions for the simulation of moving particles.

The \emph{moving\_obstacle} module facilitates simulations of fluid-particle interactions with the momentum exchange method
by implementing kernels for moving obstacle sweeps and providing the corresponding data structures.
This module furthermore provides setup functions for initializing and connecting the \pe to \walberla.
For the moving obstacle handling, an obstacle-cell relation field is provided that stores for each lattice cell 
overlapped by a \pe object a pointer to this object.
Moreover, from the \emph{lbm} module the PDF source field is accessed for the hydrodynamic force computation and the reconstruction of PDFs.
In the \emph{lbm} velocity field, body velocities are stored and accessed in the moving boundary treatment of the LBM.

For the computation of the electric potential distribution, the \emph{lse\_solver} module described in \Sect{Sec:SORMovingBCNonlinEq} is employed.
This module has been implemented for solving large sparse linear systems of equations
as they arise from the discretization of the electric potential equations (see \Sect{SubSec:FV_PBE}).
The corresponding BC classes are implemented in the \emph{pot\_bc} module described in \Sect{SubSec:BCHandlSolverModule}.
The data structures defined in the \emph{lse\_solver} module, accessed by the application and other modules,
include the stencil field representing the system matrix as well as the scalar fields for the solution and for the RHS.

The \emph{electrokin\_flow} module was designed for facilitating electrokinetic flow simulations.
This module provides kernels for coupling the involved methods 
as well as the setup and parameterization of simulations such as electrophoresis.
The setup includes initializing the stencils and RHS from the \emph{lse\_solver} module
according to the finite volume discretization of the Debye-H{\"u}ckel equation presented in \Sect{SubSec:FV_PBE}.
The kernels for imposing the electric potential BCs on the moving particles
and for computing the electrostatic forces on fluid and particles
are described in \Sect{SubSec:ElPhorSurfBCs} and \Sect{SubSec:ElPhorForceComp}, respectively.
Finally, the coupled algorithm for electrophoresis provided by the \emph{electrokin\_flow} module is presented in \Sect{SubSec:EPAlgorithm}.

\subsection{Solver for Linear Elliptic PDEs with Moving Boundaries \label{Sec:SORMovingBCNonlinEq}}
The solver module \emph{lse\_solver} in \walberla for large linear systems of equations 
has been designed as an efficient and robust black-box solver on block-structured grids~\cite{Bartuschat:2012:parallel,Bartuschat:2014:CP,Bartuschat:2016:Diss}.
To specify the problem to be solved, the application sets up the system matrix, right-hand side, and BC handling.

\begin{figure}[h!]
  \centering
      \includegraphics[scale=0.72]{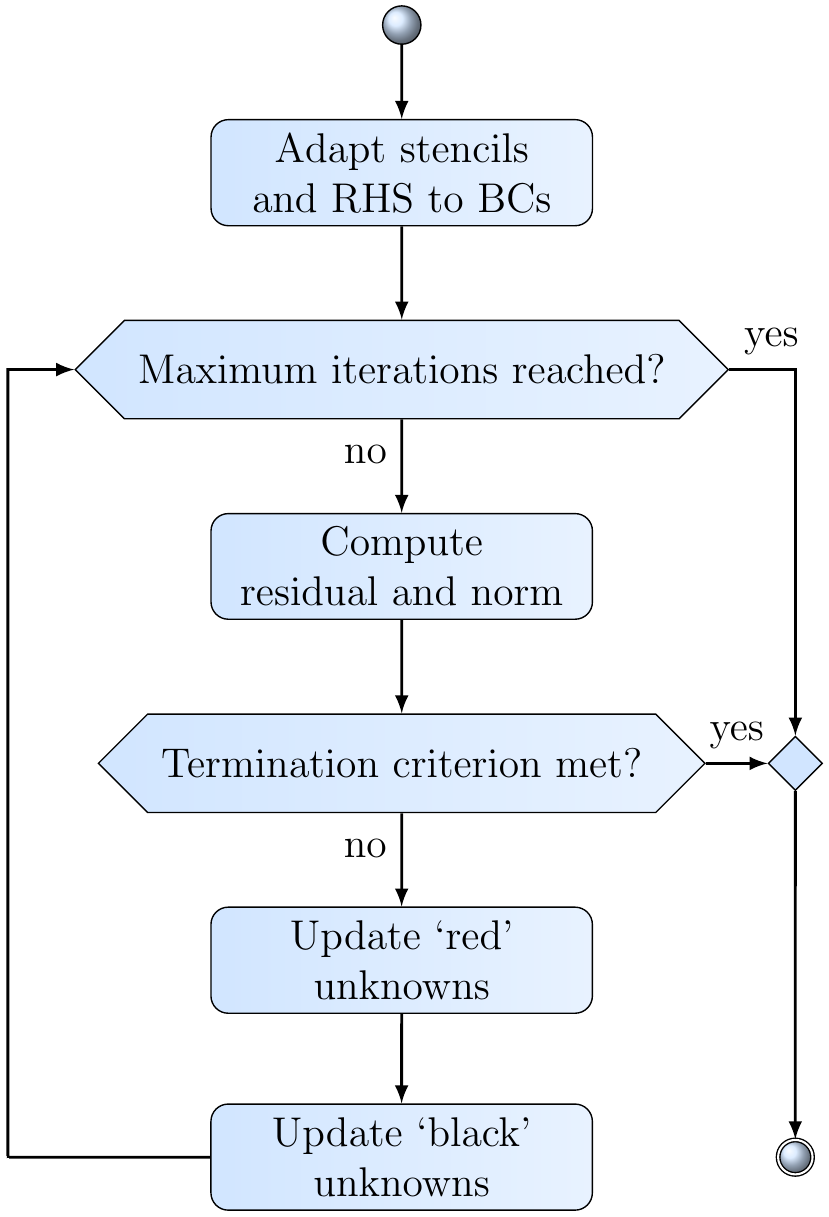}
      \captionsetup{width=0.8\textwidth}
      \caption{Activity diagram for SOR sweep \lstverb!solveTimeVaryingBCSOR! with time varying boundary conditions (BCs), \eg, due to moving particles
      \label{fig:solveTimeVaryingBCSOR}}
\end{figure}

In the \emph{lse\_solver} module, solver sweeps are pre-defined that perform the iterations at the position where they are added to the timeloop. 
For a given simulation setup, the employed solver is selected via the input file where also the solver parameters and BCs are specified.
An iterative solver requires a nested sweep that is executed until a specific convergence criterion is satisfied.
For all implemented solvers, these sweeps share a common structure. 
This structure is displayed for the SOR sweep \lstverb!solveTimeVaryingBCSOR! for moving boundaries and linear PDEs such as the Debye-H{\"u}ckel approximation in \Fig{fig:solveTimeVaryingBCSOR} as an activity diagram.
For the employed discretization of the electric potential equations,
the system matrix is represented by D3Q7 stencils.
These stencils comprise an entry for each, the center and the six cardinal directions.

In the setup function for this SOR sweep, the communication for the ghost layer exchange of the solution field in the initialization phase is set up first.
Then the SOR solver sweep is added to the timeloop,
and the kernels for relaxation, communication, and BC treatment are specified as solver sub-sweeps.
For parallel execution, the SOR algorithm is implemented in red-black order.

The filled circle at the top of the diagram in \Fig{fig:solveTimeVaryingBCSOR} indicates the starting point of the sweep in the timeloop.
At the beginning of the sweep \lstverb!solveTimeVaryingBCSOR! for moving boundaries, 
the stencils are constructed to incorporate the BCs according to the present boundary locations.
Furthermore, the RHS is adapted to these BCs before the solver iterations start.
For this purpose, a sub-sweep with a kernel for re-setting the stencils and RHS is executed before the iteration loop,
followed by the BC treatment functions \lstverb!adaptStencilsBC! and \lstverb!adaptRHSBC! described in \Sect{SubSec:BCHandlSolverModule}.
Then the standard parallel Red-Black SOR sweep is performed until the termination criterion is met.
This sweep comprises a sub-sweep for computing the residual and its $L_2$-norm for the termination criterion,
as well as two solver sub-sweeps for the SOR update of the `red' and the `black' unknowns, respectively.
In these sub-sweeps, the quasi-constant stencil optimization technique introduced in \cite{Bartuschat:2014:CP} is employed.
Based on the residual $L_2$-norm, the termination criterion for the simulations performed in this article is provided as residual reduction factor 
$\text{RES}_\text{RF}$ w.r.t. the initial norm of the simulation. 

For multigrid solvers as applied in \cite{Bartuschat:2014:CP} to charged particle simulations in absence of ions in the fluid,
the red-black update sub-sweeps in \Fig{fig:solveTimeVaryingBCSOR} are replaced by solver sub-sweeps of a V-cycle.
To apply the SOR sweep to varying stencils of linearized PDEs, such as the symmetric Poisson-Boltzmann equation,
only the sub-sweep for adapting the stencils and RHS to the BCs 
is performed at the beginning of each iteration instead of once before the iterations start (see \cite{Bartuschat:2016:Diss}).
 
\subsection{Boundary Condition Handling for Solver Module \label{SubSec:BCHandlSolverModule}}
The implicit BC handling used and initiated by the solver module has been introduced in \cite{Bartuschat:2014:CP}.
This boundary handling is based on incorporating the BCs into the stencils and right-hand side of the finite volume discretization.
That way, at an iterative update of a near-boundary value, the method implicitly uses the new values for the BCs.
For Dirichlet boundaries, the boundary values are linearly extrapolated to the boundary cell and for Neumann BCs the boundary values are approximated by central differences.
For both the stencils and the right-hand side, a direction-dependent BC treatment is used.

The functions for this BC treatment are implemented in the \emph{pot\_bc} module.
This module employs own \emph{nonBC} and \emph{nearBC} flags for the BC handling of scalar potentials.
Moreover, for each BC class in this module an associated \emph{BC} flag is defined.
For the employed cell-centered discretization, the module contains one class for each, Neumann and Dirichlet BCs.

For incorporating the BCs into the stencils the kernel \lstverb!adaptStencilsBC! is implemented.
This kernel iterates over all lattice cells to find scalar potential \emph{nearBC} cells.
At each cell with \emph{nearBC} flag, the kernel employs the D3Q7 stencil directions
to iterate over the neighboring cells. 
In directions of a cell with scalar potential \emph{BC} flag, the stencil entry of the \emph{nearBC} cell, 
associated with the direction of the \emph{BC} flag, is adapted accordingly.

The function \lstverb!adaptRHSBC! 
employs the standard boundary handling of the \emph{bc} module 
to invoke the \emph{pot\_bc} kernels for adapting the RHS to the BCs.
To this end, the direction-dependent BC treatment kernels in the corresponding BC classes 
implement the RHS adaption depending on the BC value.
The BC value is specified in the input file for static BCs,
or in a previous time step for BCs of moving particles (see \Sect{SubSec:ElPhorSurfBCs}).
To facilitate such complex boundaries, the BC classes store the BC values and the corresponding boundary cell ranges. 
The latter are stored in a memory-efficient way either as cell intervals or as cell sets.
Moreover, to allow the computation of scalar potential gradients directly from the solution field,
the BC values are set in this field at boundary cells when the RHS is adapted.
From these BC values and from the solution at the \emph{nearBC} cell, the value at the boundary cell required for the gradient can be extrapolated.

\subsection{Electric Potential Boundary Condition Handling for Moving Particles \label{SubSec:ElPhorSurfBCs}}
Prior to the EDL potential distribution computation by the \emph{lse\_solver} module,
uniform $\zeta$-potentials or surface charge densities
are imposed at the moving particles by means of scalar potential BCs.
These electrical surface properties are specified in the input file for different particle types with a common \emph{uid} (unique identifier) defined in the \pe.
To this end, the sweep function \lstverb!setPotBC_ChargParticles!
is implemented in the \emph{electrokin\_flow} module that 
maps the charged particles onto the lattice and sets the electric potential BC values and the associated BC handling flags at the corresponding cells.
The function first overwrites the values of all cells in the RHS field with zero  to remove the values from the previous BC treatment. 
Then the mapping is performed for all movable rigid bodies located in the subdomain of each process.
For each particle the mapping is conducted in an extended axis-aligned bounding box that surrounds this rigid body and the associated \emph{nearBC} flags.
The mapping is realized in three steps:
\begin{enumerate}
   \item[1)]First, all scalar potential \emph{nearBC} and \emph{BC} flags from the previous time step are removed, 
            and the scalar potential \emph{nonBC} flags are set.
            Moreover, the previous BC values and the associated cells in the BC class instances are removed.
   \item[2)]Then, for particles with prescribed electric BCs,
            the BC handling for the \emph{lse\_solver} module (see \Sect{SubSec:BCHandlSolverModule}) is prepared.
            For each rigid body with a \emph{uid} for which a surface property is specified in the input file,
            the associated BC is obtained.
            The cells overlapped by this particle are gathered and are added together with the BC value to the corresponding BC class instance.
            Moreover, the \emph{BC} flag is set at the overlapped cells.
   \item[3)]Finally, the \emph{nearBC} flag is set at all cells adjacent to a \emph{BC} cell. 
\end{enumerate}
Each step is performed for all bodies on a process before the next step begins,
to prevent that for particles with overlapping bounding boxes
the flags from a previous step are overwritten.

\subsection{Computing Electrostatic Forces on Fluid and Particles \label{SubSec:ElPhorForceComp}}
The electric forces acting on the ions in the fluid are incorporated into the incompressible Navier-Stokes equation by 
the body force term $\vec{f}_b = -\rho_e(\psi) \, \gradOp \left( \varphi + \psi \right)$,
which coincides with the corresponding term employed in Henry's solution (see \Sect{Sec:ElkinFlow}).
Due to the linear superposition of the electric potential components, the gradient is applied to both components separately.
Since the applied field $\vec{E}_\text{ext}=-\gradOp \varphi$ is given,
only the EDL potential gradient must be computed.

For the computation of the electric field due to the EDL, the \emph{electrokin\_flow} module
provides a kernel that performs the gradient computation at all scalar potential \emph{nonBC} cells.
The gradient of the electric potential is computed, 
as previously introduced in~\cite{Bartuschat:2014:CP}, 
by means of finite differences that provide $\mathcal{O}({\textup{dx}}^2)$ accuracy.
Where possible, an isotropy-preserving D3Q19 stencil is used (\cf{} \cite{2013:Ramadugu:LatticeDiffOp}) instead of a D3Q7 stencil.
With the LBM D3Q19 stencil, the gradient can be computed using $w_q$-weighted differences of neighboring values in 18 directions $\vec{e}_q$ as
\begin{equation}
   \gradOp \psi(\vec{x}_b) \approx \frac{1}{w_1} \sum\limits_{q=2}^{19} w_q \, \psi(\vec{x}_b + \vec{e}_q ) \cdot \frac{\vec{e}_q }{\textup{dx}^2}.
    \label{Eq:ElPotGradComp}
\end{equation}
At \emph{nearBC} cells the D3Q7 stencil is applied 
to compute the gradient of $\psi$ from the BC values stored at particle and boundary cells in the BC treatment (see \Sect{SubSec:BCHandlSolverModule}).
The obtained electric field is stored in a field of cells that is accessed in the body force computation.

For the computation of the electric body force 
and of the electrostatic force exerted on the particles,
a further kernel is implemented in the \emph{electrokin\_flow} module:\\
The kernel first iterates on each parallel process 
over all lattice cells to compute the body force at scalar potential \emph{nonBC} cells.
This force is computed as product of charge density $\rho_e(\psi)$ and 
total electric field $\vec{E}_\text{total} = \vec{E}_\text{ext} -\gradOp \psi$.
The relation of charge density and EDL potential follows \Eqn{Eq:ChrgDensDHA}.
The obtained electric body force is written to the external force field of the \emph{lbm} module that is accessed by the LBM kernels with forcing.
Then the kernel iterates over all non-fixed particles residing on 
the current parallel process to compute the electrostatic force.
For each of these particles the force is computed 
from the particle charge and the applied field as $\vec{F}_C=q_e\,\vec{E}_\text{ext}$
and is then added to the particle.

\subsection{Algorithm for Electrophoresis \label{SubSec:EPAlgorithm}}
The overall parallel algorithm for electrophoresis simulations with \walberla is shown in \Alg{Alg:ElPh_mine}.
After the setup and initialization phase, the electric BCs for the EDL potential are set at the moving charged particles by means of \lstverb!setPotBC_ChargParticles! at each time step.
Then the Debye-H{\"u}ckel approximation is solved by means of the SOR sweep \lstverb!solveTimeVaryingBCSOR!.
The iterations are performed until the specified termination criterion for the residual is met.
\begin{algorithm}[h!]
   \caption{Electrophoresis Algorithm}
   \ForEach{time step, }
   {
      {\color{blue} // solve Debye-H{\"u}ckel approximation (DHA): }   \\
      set electric BCs of particles                                        \\
      \While{residual $\geq$ tol}
      {
         apply SOR iteration to DHA                               \\
      }
     
      {\color{blue} // couple potential solver and LBM:}          \\
      \Begin {
      compute electric field due to EDL                           \\ 
      compute charge density in fluid                             \\ 
      compute electric body force                                 \\
      apply electrostatic force to particles                      \\ 
      }
      {\color{blue} // solve lattice Boltzmann equation with forcing, considering particle velocities:}\\
      set velocity BCs of particles                                        \\
      \Begin {
         perform fused stream-collide step \\
      }
      {\color{blue} // couple potential solver and LBM to \pe{}: } \\
      \Begin {
      apply hydrodynamic force to particles                       \\
      \pe\ moves particles depending on forces                    \\
      }
   }
   \label{Alg:ElPh_mine}
\end{algorithm}

From the obtained EDL potential distribution and the applied field the electric body force exerted on the fluid is computed,
as described in \Sect{SubSec:ElPhorForceComp}.
First the kernel computing the electric field caused by the EDL is applied.
Then the kernel for the electric body force computation from the charge density distribution in the fluid 
and the total electric field is invoked.
This kernel additionally applies the electrostatic force exerted by the applied field to the particles.

Then the rigid body mapping sweep described in \Sect{SubSec:MEA} is performed,
imposing the particle velocities for the subsequent LBM sweep.
In that parallel sweep, an LBM kernel with fused stream-collide step and forcing term (see \Sect{SubSec:TRT}) is employed to compute the fluid motion influenced by the moving particles and by the electrostatic force exerted on ions in the EDL.

After the LBM sweep, the hydrodynamic forces on the particles are computed by the momentum exchange method. 
The obtained hydrodynamic force contributions and the electrostatic forces are then aggregated by the \pe{}.
From the resulting forces and torques, the new particle velocities and positions are computed in the subsequent \pe{} simulation step
by the PFFD algorithm~\cite{iglberger:2010:Pe} that additionally resolves rigid body collisions.%
\section{Electrophoresis Simulations \label{Sec:ElPhorResults}}
In the following, the correctness of the electrophoresis algorithm is validated and its parallel performance is analyzed.
The electric potential computation is validated for a sphere with uniform surface charge surrounded by an EDL.
This sphere is placed in a micro-channel subject to an applied electric field.
Moreover the flow field caused by the electrophoretic motion of the sphere in the micro-channel is visualized, 
together with the electric potential and the surrounding ions to qualitatively show the correctness of the simulations.
Then the electrophoretic velocity and the retardation by the counter-ions in the EDL 
is validated w.r.t.\ Henry's solution for different sphere sizes and values of $\kappa \, R$.
Finally, weak scaling experiments are presented and the parallel performance and scalability of the electrophoresis algorithm 
is shown up to more than four million spheres.

\subsection{Simulation Setups  \label{SubSec:SimSetup}}
In all simulations, the insulating spherical particles are suspended in a symmetric aqueous electrolyte solution with the physical parameters listed in  \Tab{Tab:Electrolyte_Sphere_Parameters}.
\begin{table}[h!]
   \begin{center}
      \caption{Parameters of electrolyte solution and particles used in all simulations.
                    Shown are kinematic viscosity $\nu_f$, density $\rho_f$, permittivity $\varepsilon_e$, ion valence $z$, temperature $T$, and particle density $\rho_p$. \label{Tab:Electrolyte_Sphere_Parameters}}
      \begin{tabular}{ccccc|c}
\toprule
      $\nu_f $                                                            &  $\rho_f$                                                       & $\varepsilon_e$                        &  $z$        & $T$                         &  $\rho_p$    \\
\midrule
      $\SI{1.00e-6}{\square\meter\per\second}$      & $\SI{1000}{\kilogram\per\cubic\meter}$    & $78.54 \cdot \varepsilon_0$     &  $1$        & $\SI{293}{\kelvin}$  & $\SI{1195}{\kilogram\per\cubic\meter}$     \\
\bottomrule
      \end{tabular}
   \end{center} 
\end{table}
Moreover, gravitational are effects neglected to ensure that the particle motion is driven solely by electric forces.
The EDL thickness $\lambda_D$ is in the order of the particle diameter in all simulations
and always greater than the $12$ lattice sites required for a sufficient resolution,
as observed in the electro-osmotic flow simulations in~\cite{Bartuschat:2016:Diss}.
A uniform surface-potential distribution is chosen with a $\zeta$-potential sufficiently low to approximate the Poisson-Boltzmann equation by the Debye-H{\"u}ckel approximation.
For this approximation analytical solutions are known for the potential distribution and for the electrophoretic velocity.
The linear system of equations resulting from the finite volume discretization of the Debye-H{\"u}ckel equation is solved by the SOR method
that is sufficient for the quickly decaying electric potential due to the counter-ions in the EDL.
For the SOR, a relaxation parameter of $\omega_\text{SOR}=1.7$ is applied in all simulations.
The LBM is employed with TRT operator and second-order forcing term by Luo and re-defined momentum density (see \Sect{SubSec:TRT}).

\begin{figure}[h]
  \centering
    \includegraphics[width=0.6\textwidth]{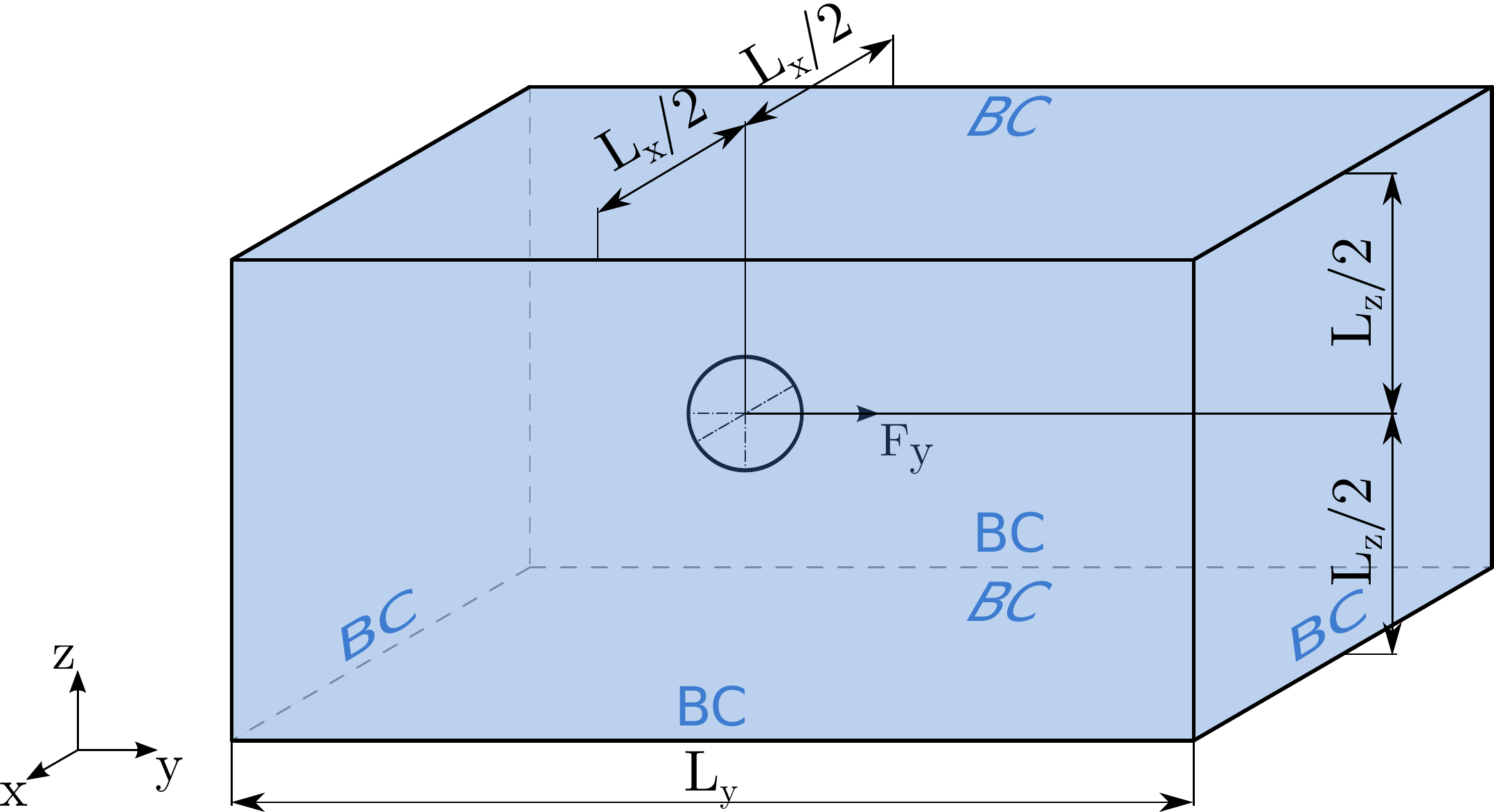}
      \caption[Electrophoresis validation setup.]{Setup for electrophoresis of spheres in square duct with different BCs.}
  \label{fig:WallEffectSetup}
\end{figure}

For the validation simulations the setup with a single sphere is depicted in \Fig{fig:WallEffectSetup}.
The sphere is placed on the longitudinal axis of a cuboid domain of size $L_\text{x} \times L_\text{y} \times L_\text{z}$ at an initial position of $\text{y}_\text{0}$,
and an electrostatic force in y-direction acts on the sphere.

For the EDL potential validation in \Sect{SubSec:ElectrophorPot} the parameters in the left part of  \Tab{Tab:EDL_ElectrophorMotion_Parameters} are used.
The simulation domain is discretized with the lattice spacing $\textup{dx}$ and at the surface of the sphere of radius $R_L$ the $\zeta$-potential is imposed.
In addition to the parameters in \Tab{Tab:Electrolyte_Sphere_Parameters} the bulk ion concentration $c^{\infty}$ is simulated 
to obtain the displayed values of the Debye-H{\"u}ckel parameter $\kappa$ and of the characteristic EDL thickness $\lambda_D$.\\
\begin{table}[h!]
   \begin{center}
      \caption{Simulation parameters for validation of EDL potential (left) and electrophoresis of sphere in micro-channel  (whole table). \label{Tab:EDL_ElectrophorMotion_Parameters}}
      \begin{tabular}{ccc|cc}
\toprule
      $\zeta $                             &  $c^\infty $                                & $\kappa $                             &  $q_s $                                                    & $E_\text{y}$     \\
\midrule
      $\SI{-10.0}{\milli\volt}$       & $\SI{5.00e-6}{\mol\per\litre}$   & $\SI{7.41e6}{\per\meter}$     &  $\SI{-19.9e-18}{\ampere\second}$        & $\SI{-4.7e6}{\volt\per\meter}$       \\
\midrule
\midrule
     $\textup{dx} $                     &  $R_L$                                  &  $\lambda_{D,L}$                     & $\tau$                                                     & $\textup{dt} $   \\ 
\midrule
      $\SI{10e-9}{\meter}$         & $12$                                     &  $\num{13.49}$                         & $\num{6.5}$                                            & $\SI{200e-12}{\second}$     \\ 
\bottomrule
      \end{tabular}
   \end{center} 
\end{table}
For the electrophoretic motion visualization in \Sect{SubSec:ElectrophorVelMicrochan}
additionally the parameters in the right part of \Tab{Tab:EDL_ElectrophorMotion_Parameters} are employed.
The sphere's charge $q_s$ is obtained from the surface charge density $\sigma_s$ according to the $\zeta$--$\sigma_s$ relationship~\eqref{Eq:SurfChrgZetaRelationOhshima},
multiplied by the surface area of the sphere, as $q_s = 4 \pi R^2 \sigma_s$~\cite{Hunter:1981:ZetaPotCollSci}.
The high electric field magnitude $E_\text{y}$ is chosen to keep the number of simulation time steps at a minimum.
From the applied LBM relaxation parameter $\tau$ the time increment $\textup{dt}$ results according to \Eqn{Eq:kinVisc} from the chosen viscosity $\nu_f$ and $\textup{dx}$.

The velocity validation experiments in \Sect{SubSec:ElectrophorVel_ValidOpenDom} and scaling experiments in \Sect{SubSec:ParallelPerf}
are performed on SuperMUC\footnote{\url{www.lrz.de/services/compute/supermuc/}} of the Leibniz Supercomputing Centre LRZ in Garching (Germany).
For these experiments the parameters in \Tab{Tab:ElectrophorValidSpheres_Parameters} are used.
The $\zeta$-potential has the same absolute value as in the electrophoretic motion simulations in \Sect{SubSec:ElectrophorVelMicrochan}.
From the value of $c^{\infty}$ and from the electrolyte parameters in \Tab{Tab:Electrolyte_Sphere_Parameters} results the value of $\kappa$ 
and a Debye length $\lambda_D$ of approximately 15 lattice sites.
The LBM is employed with the relaxation time $\tau=6$, yielding the time increment $\textup{dt}$.
\begin{table}[h!]
   \begin{center}
      \caption{Simulation parameters for electrophoretic velocity validation w.r.t.\ Henry's solution and for scaling experiments.
                    Parameters with subscript 'Sc' are applied in scaling runs. \label{Tab:ElectrophorValidSpheres_Parameters} }
      \begin{tabular}{cccc|c}
\toprule
      $\zeta $                            &  $c^\infty $                              & $\kappa $                                &  $E_\text{y}$                                      & $E_\text{y,Sc}$                              \\
\midrule
      $\SI{10.0}{\milli\volt}$      & $\SI{1.60e-5}{\mol\per\litre}$   & $\SI{13.3e6}{\per\meter}$       & $\SI{99.0e6}{\volt\per\meter}$            & $\SI{38.0e6}{\volt\per\meter}$     \\
\midrule
\midrule
     $\textup{dx} $                    & $\lambda_{D,L}$                    & $\tau$                                      & $\textup{dt} $                                      &  $R_\text{L,Sc}$                            \\ 
\midrule
      $\SI{5.00e-9}{\meter}$      &  $\num{15.08}$                      & $\num{6}$                               & $\SI{45.8e-12}{\second}$                    &  $6$                                             \\ 
\bottomrule
      \end{tabular}
   \end{center} 
\end{table}
As in \Sect{SubSec:ElectrophorVelMicrochan}, a high electric field $E_\text{y}$ is chosen for the validation to keep the number of simulation time steps low.
In the scaling experiments the electric field is reduced to a more realistic magnitude of $E_\text{y,Sc}$ and spheres of radius $R_\text{L,Sc}$ are simulated.
To obtain different values of $\kappa \, R$, the sphere radii in \Tab{Tab:ElectrophoresisValidSpheres_Params} are used in the validation experiments.
\begin{table*}[h]
   \begin{center}
      \caption[]{Surface charges $q_s$ used in velocity validation (and scaling) experiments.\label{Tab:ElectrophoresisValidSpheres_Params}}
      \begin{tabular}{l|cccccccc}
         \toprule
         $R_L$                                       &   4       &   6       &   8       &   9       &   12      \\
         \midrule
         $q_s / (\SI{e-18}{\ampere\second})$         & 2.21      &  3.67     & 5.36      & 6.29      & 9.43      \\
         \bottomrule
      \end{tabular}
   \end{center}
\end{table*}
For the sphere sizes and parameters used in the scaling and validation experiments, the associated surface charges are displayed in \Tab{Tab:ElectrophoresisValidSpheres_Params}.
These charges are again obtained from the general $\zeta$--$\sigma_s$ relationship~\eqref{Eq:SurfChrgZetaRelationOhshima}.
For all parameters in \Tab{Tab:ElectrophorValidSpheres_Parameters},
the maximum relative deviation of $\sigma_s$ for the general relationship from the exact value for the Debye-H{\"u}ckel approximation is below 0.2\%.

\subsection{Electric Potential in an EDL around a Sphere \label{SubSec:ElectrophorPot}}
To validate the computation of the EDL potential $\psi$ around a charged particle,
a sphere is simulated in a large domain of size $128 \, \textup{dx} \times{} 256 \, \textup{dx} \times{} 128 \, \textup{dx}$.
The analytical solution of the Debye-H{\"u}ckel equation~\eqref{Eq:DH_Radial}
representing the EDL potential around the sphere is given by \Eqn{Eq:PotDistrSphereDH}.
For the validation, a spherical particle of radius $R_L=12$ with initial position 
$\text{y}_\text{0} = 64 \, \textup{dx}$ is chosen.

Despite its quick decay, the analytical solution of the electric potential at the domain boundaries differs from zero. 
Thus, the values of $\psi$ at these boundaries are set to the analytical solution by means of Dirichlet conditions.
To solve the Debye-H{\"u}ckel equation subject to these BCs,
a residual reduction factor of ${\text{RES}_\text{RF} = \num{2e-7}}$ is employed as termination criterion for the SOR method.

\begin{figure}[h]
   \centering
        \includegraphics[scale=1.05]{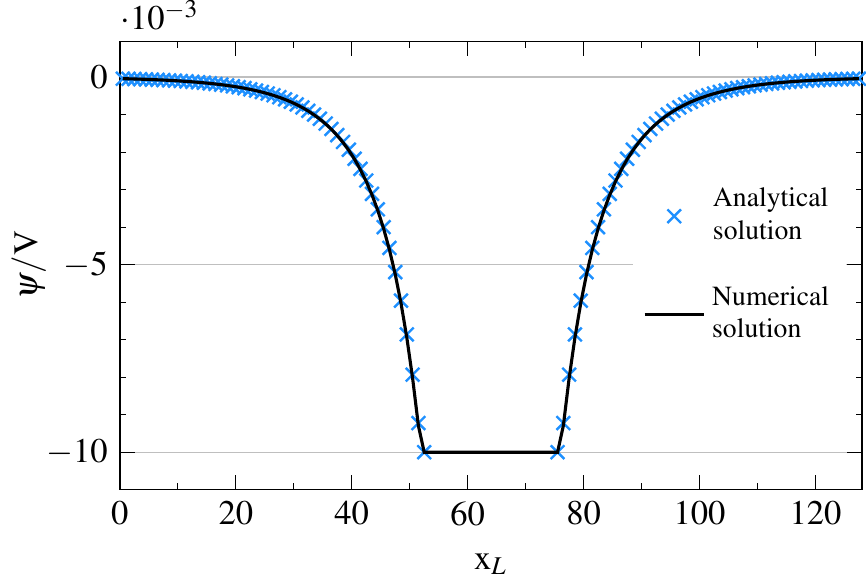}
      \caption[Analytical and numerical solution for EDL potential of charged sphere]{Analytical and numerical solution for EDL potential of sphere with uniform surface charge. \label{fig:CP_EOF_Electrophor_ElPotValidation} }
\end{figure}

The analytical ($\psi$) and numerical ($\psi^*$) solution at the initial particle position are depicted in \Fig{fig:CP_EOF_Electrophor_ElPotValidation} 
along a line in x-direction through the sphere center.
Both graphs agree very well, showing the correctness of
the finite volume discretization and the applied SOR solver,
as well as the boundary handling at the particle surface.
Inside the insulating particle, the electric potential is not computed explicitly, due to the uniform surface potential and the resulting symmetric distribution of $\psi$ in the sphere.
The electrostatic force needed to compute the particle motion is computed directly from the applied field and the particle charge, instead of the electric potential gradient as in \cite{Bartuschat:2014:CP}.

\subsection{Electrophoresis of a Sphere in a Micro-Channel \label{SubSec:ElectrophorVelMicrochan}}
The application of the electric field in \Tab{Tab:EDL_ElectrophorMotion_Parameters} to the micro-channel setup described in \Sect{SubSec:SimSetup}
gives rise to an electrophoretic motion of the sphere.
Due to the applied field and the resulting electrostatic force of $F_\text{C,y}=\SI{933e-12}{\newton}$ 
the particle moves in y-direction, retarded by the channel walls and the opposing force on the EDL.
For the chosen parameters, the terminal particle speed of $U_\text{EP}=\SI{224.5}{\milli\meter\per\second}$  (or $U_{\text{EP},L} = \num{4.49e-3}$)
is obtained for free space according to Henry's solution~\eqref{Eq:ElPhorVelSphere},
corresponding to a particle Reynolds number of $\textup{Re}_{p,d} = 0.054$.
Since gravitational effects are neglected, the particle density in \Tab{Tab:Electrolyte_Sphere_Parameters} only has an impact on the time required to reach steady-state.

In the simulation, periodic BCs are applied in y-direction 
of the domain of size $128 \, \textup{dx} \times{} 256 \, \textup{dx} \times{} 128 \, \textup{dx}$.
At all other walls, no-slip conditions are applied for the LBM and homogeneous Neumann conditions for the electric potential.
At each time step, the Debye-H{\"u}ckel equation is solved by SOR with the termination criterion ${\text{RES}_\text{RF} = \num{2e-7}}$.

\begin{figure*}[h!]
   \centering
   \subfigure[Results after \num{5001} time steps. \label{fig:CP_EOF_Electrophor_a}]{  
      \includegraphics[trim = 0mm 36mm 0mm 37mm, clip, width=0.83\textwidth,keepaspectratio ]{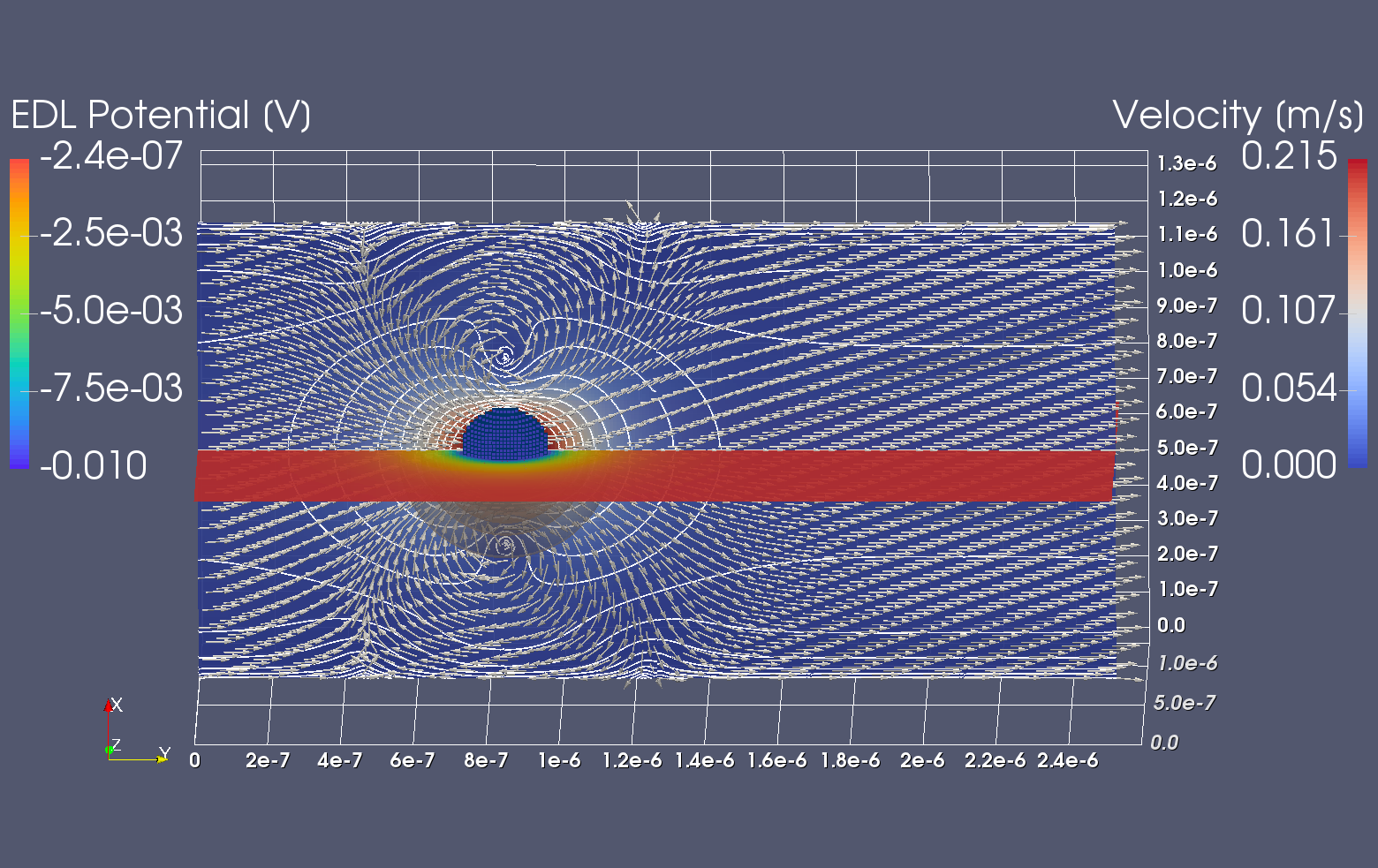}
   }\\  
   \subfigure[Results after \num{30001} time steps. \label{fig:CP_EOF_Electrophor_b}]{ 
      \includegraphics[trim = 0mm 36mm 0mm 37mm, clip, width=0.83\textwidth,keepaspectratio]{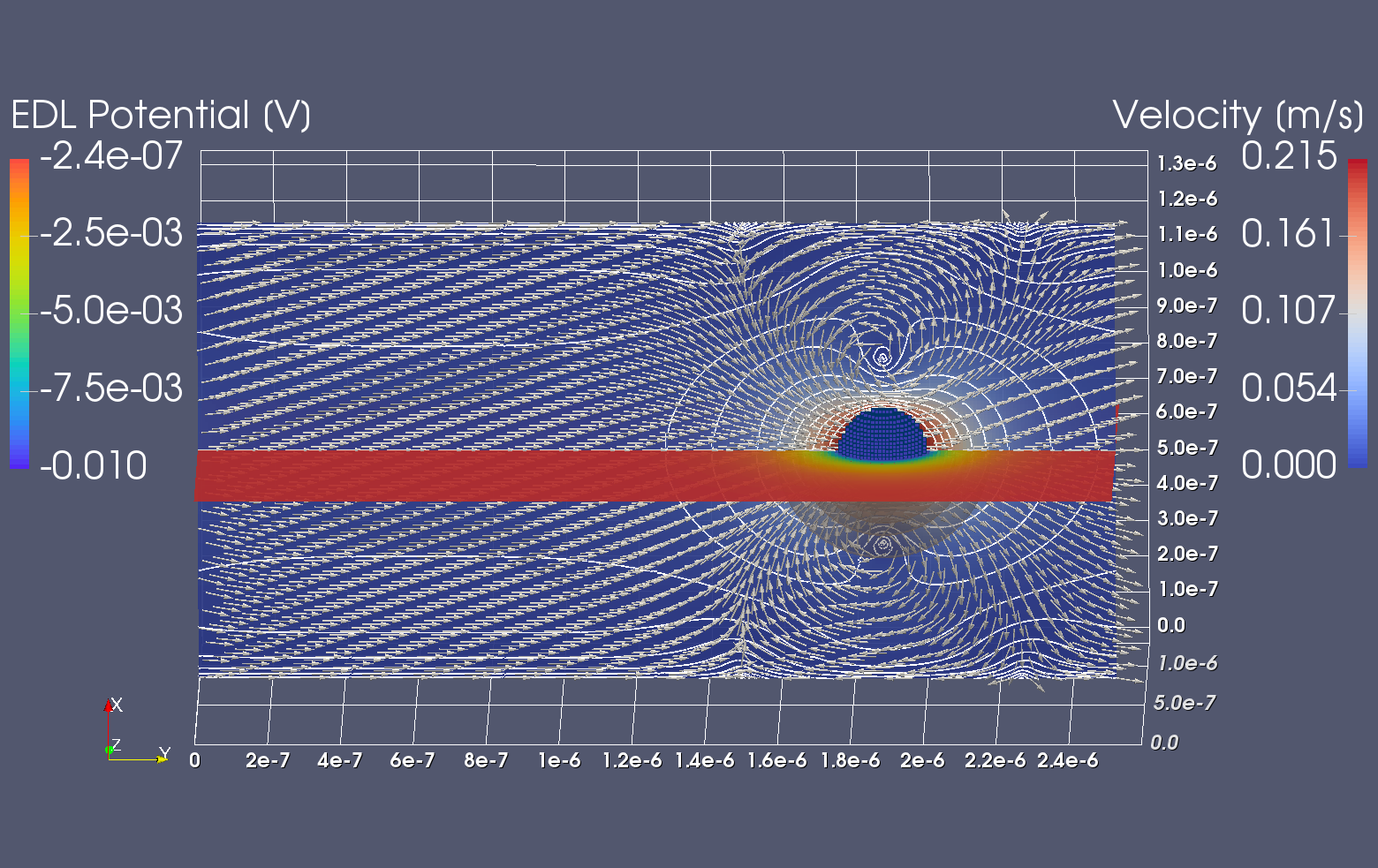}
   }
  \caption[Electrophoresis - flow field, potential and ion distribution around charged particle.]{Electrophoresis of spherical particle in micro-channel with insulating no-slip walls. 
                                          Visualization of flow field in x-y plane, EDL potential in y-z plane, and ion charge distribution (as equipotential surfaces) around charged particle. \label{fig:CP_EOF_Electrophor} }
\end{figure*}

In \Fig{fig:CP_EOF_Electrophor}, the results of the electrophoresis simulation are visualized at different time steps.
The EDL potential $\psi$ in the y-z plane through the domain center is displayed,
together with semi-transparent hemispherical equipotential surfaces representing the excess counter-ions in the EDL.
The flow field around the moving sphere is visualized in the x-z plane through the domain center.
Arrows of uniform length indicate the flow direction, while the velocity magnitude is represented by the shown color-scale and 
by twelve white isosurface contour lines with logarithmic intervals in the range of \SI{13.67e-3}{\meter\per\second} to \SI{6.08e-6}{\meter\per\second}.

The flow field around the moving particle shown in \Fig{fig:CP_EOF_Electrophor_a} is fully developed after \num{5001} time steps, 
and the particle has already attained its terminal velocity.
Due to the periodicity of the domain, a channel flow in axial direction has emerged from the particle motion, as indicated by the contour lines.
Moreover, a vortex has formed between the sphere and the surrounding no-slip walls.
The flow field moves with the particle that translates along the channel centerline (see \Fig{fig:CP_EOF_Electrophor_b}).
Because of the equilibrium representation of the EDL, the distribution of $\psi$ is at all time steps symmetric w.r.t.\ the sphere center, almost up to the boundary.

\subsection{Validation of Electrophoretic Motion of a Sphere \label{SubSec:ElectrophorVel_ValidOpenDom}}
To quantitatively validate the overall electrophoresis algorithm,
the electrophoretic velocity of spheres with uniform surface charge
is compared to Henry's solution~\eqref{Eq:ElPhorVelSphere} for a spherical particle in an unbounded electrolyte solution.
In the simulation experiments, a sphere with the $\zeta$-potential in \Tab{Tab:ElectrophorValidSpheres_Parameters}
is moving under the influence of an applied electric field in a large domain filled with an electrolyte solution.
The simulations are performed for different values of $\kappa \, R$ by varying the sphere radius while keeping the EDL thickness constant.
For the different sphere radii, the forces $F_\text{C}$ on the particle resulting from the applied field in \Tab{Tab:ElectrophorValidSpheres_Parameters} 
and the associated surface charges, are displayed in \Tab{Tab:ElectrophoresisValidSpheres_DomainSizes}.
For validation, the relative deviation $\Delta_\text{r}U$ of the obtained terminal sphere velocity from the theoretical velocity in \Eqn{Eq:ElPhorVelSphere} is evaluated.
The employed simulation parameters given in \Sect{SubSec:SimSetup} are chosen such that the electrophoretic motion is in the Stokes regime
and the EDL potential is governed by the Debye-H{\"u}ckel approximation.
Therefore, the superposition principle is assumed to hold.
Thus, from the obtained relative deviation from the analytical solution in an unbounded domain $\Delta_\text{r}U$,
the relative deviation due to wall effects and volume mapping errors $\Delta_\text{r}U_\text{Stokes}$ will be subtracted.
These relative deviations were examined in \cite{Bartuschat:2016:Diss} for several domain sizes and sphere radii.
For no-slip BCs, the wall effect was shown to comply with analytical and experimental results.
In the experiments, domain sizes are used for which the relative deviation from Stokes velocity is close to $\Delta_\text{r}U_\text{Stokes}=\SI{-3}{\percent}$ for free-slip BCs.

For the different sphere radii and the associated values of $\kappa \, R$, the electrophoretic velocities 
according to Henry's solution are displayed in \Tab{Tab:ElectrophoresisValidSpheres_DomainSizes}.
These velocities correspond to particle Reynolds numbers $\textup{Re}_{p,d}$
from $0.018$ to $0.057$ for the particle diameters of \SI{40}{\nano\meter}
to \SI{120}{\nano\meter}. \\
To quantify the retardation by the opposing force on the EDL,
the variable $\text{EP}_\text{Ret} = \frac{U_\text{EP} - U_\text{EM}}{U_\text{EM}}$ is introduced.
This variable represents the relative deviation
of the electrophoretic velocity of a particle with charge $q_s$ in presence of the electric double layer
from the migration velocity $U_\text{EM}$ of a particle with the same charge in absence of surrounding ions.
For the examined sphere radii, this retardation is in the range of $20\%$ to $42\%$.\\
\begin{table*}[ht]
   \begin{center}
      \caption[]{Electrophoresis parameters and domain sizes dependent on sphere radii $R_L$.
                  For the relations of sphere radius to EDL thickness $\kappa \, R$, 
                  electrostatic forces $F_\text{C}$, theoretical electrophoretic velocities $U_\text{EP}$, 
                  Reynolds numbers $\textup{Re}_{p,d}$, and electrophoretic retardation $\text{EP}_\text{Ret}$ are given.
                  Listed in lower part are domain sizes per dimension $L_{\text{x},\text{y},\text{z}}$, initial sphere positions $\text{y}_\text{0}$,
                  process numbers per dimension $\text{\#proc}_{\text{x},\text{y},\text{z}}$, and relative deviations of sphere velocities in free-slip domain 
                  from Stokes velocity $\Delta_\text{r}U_\text{Stokes}$. \label{Tab:ElectrophoresisValidSpheres_DomainSizes}}
      \begin{tabular}{l@{\hspace{0.5ex}}|c@{\hspace{1.5ex}}c@{\hspace{1.5ex}}c@{\hspace{1.5ex}}c@{\hspace{1.5ex}}c@{\hspace{1.5ex}}c@{\hspace{1.5ex}}c@{\hspace{1.5ex}}c}
         \toprule
         $R_L$                                                                   &   4       &   6       &   8       &   9       &   12      \\
         \midrule
         $\kappa \, R$                                                              & 0.265     &  0.398    & 0.530     & 0.597     & 0.796     \\
         $F_\text{C} / (\SI{e-12}{\newton})$                                     & 219       &  363      & 530       & 622       & 934       \\
         $U_\text{EP} / (\si{\milli\meter\per\second})$                           & 461       & 464       & 466       & 468       & 471       \\[1pt]
         $\textup{Re}_{p,d}$                                                     & 0.018     &  0.028    & 0.037     & 0.042     & 0.057     \\
         $\text{EP}_\text{Ret} / \si{\percent}$                                  & -20.6     &  -27.7    & -33.6     & -36.2     & -42.8     \\
         \midrule
         \midrule
         $L_{\text{x},\text{z}} / \textup{dx}$                                   &  864      &   1280    &   1632    &   1632    &   1632    \\
         $L_\text{y\phantom{,z}} / \textup{dx}$                                  &  1248     &   1536    &   2048    &   2048    &   2048    \\
         $\text{y}_\text{0} / \textup{dx}$                                       &  392      &   588     &   784     &   882     &   1176    \\
         \midrule
         $\text{\#proc}_{\text{x,y,z}}$                                          &  $8 \times 16 \times 8$ & $8 \times 16 \times 16$ & $16 \times 16 \times 32$ & $16 \times 16 \times 32$ & $16 \times 16 \times 32$ \\
         \midrule                                          
         $\Delta_\text{r}U_\text{Stokes} / \si{\percent}$                        & -3.04     &  -2.65    &  -2.98    &  -2.89    &  -3.00    \\
         \bottomrule
      \end{tabular}
   \end{center}
\end{table*}
Then the domain sizes are listed in~\Tab{Tab:ElectrophoresisValidSpheres_DomainSizes},
together with the initial positions in movement direction that correspond to $\text{y}_\text{0}=98 \times R_L$.

For the electric potential, homogeneous Neumann BCs are applied at the walls in x- and z-direction as
${\restr{\frac{\partial \Phi}{\partial \vec{n}} }{ \Gamma_W \cup \Gamma_E \cup \Gamma_B \cup \Gamma_T } = \SI{0}{\volt\per\meter}}$.
To improve convergence~\cite{Bartuschat:2014:CP}, homogeneous Dirichlet BCs 
are applied at the walls in y-direction as ${\restr{\Phi}{\Gamma_N \cup \Gamma_S } = \SI{0}{\volt}}$. 
The particle is located at sufficient distance from these walls for the EDL potential to decay to approximately zero,
and therefore these BCs do not affect the electric potential distribution.
As termination criterion of the SOR solver for the Debye-H{\"u}ckel equation, a residual reduction factor of $\text{RES}_\text{RF} = \num{1e-6}$ is used.

The parallel simulations are run on 64 to 512 nodes of SuperMUC
with the numbers of processes listed in \Tab{Tab:ElectrophoresisValidSpheres_DomainSizes}.
Within the execution time of \SI{37}{\hour} to \SI{48}{\hour}, a number of \num{70296} to \num{74000} time steps are performed,
and the spheres cover a distance of about $300\,\textup{dx}$.
The high numbers of time steps are chosen to ensure that the spheres reach steady-state motion, 
and that large numbers of sampling values are available to compute the terminal velocities.

The sphere velocities are sampled every 20 time steps during the simulation.
In the second half of the simulation, the terminal particle velocity is reached.
Thus, the mean particle velocity $U^*$ 
and the velocity fluctuations $\delta_U=\frac{U^*_\text{max}-U^*_\text{min}}{U^*}$ due to volume mapping effects 
are computed from the last 50\% output values.\\
In the considered range of time steps
additionally the number of time steps between two SOR calls
and the number of SOR iterations is monitored.
The average number of time steps between two SOR calls decreases 
from $\Delta \text{TS}_\text{SOR} = 24$ to $\Delta \text{TS}_\text{SOR} = 3$ as $R_L$ increases from $4$ to $12$.
Likewise, the average number of SOR iterations per solver call decreases 
from \num{451} iterations to \num{198} iterations for the respective sphere radii. \\
The obtained terminal velocity in lattice units $U_L^*$ and the fluctuations are displayed in \Tab{Tab:ValidElectrophoresis}.
\begin{table}[h]
   \centering
   \caption[]{Simulation results of electrophoretic velocity validation for different sphere sizes.
              Shown are the theoretical velocities $U_{\text{EP},L}$ for unbounded domains in lattice units, 
              obtained velocities $U_L^*$ and fluctuations $\delta_U$,
              relative deviations $\Delta_\text{r}U$ of $U_L^*$ from $U_{\text{EP},L}$, 
              and relative deviations $\Delta_\text{r}U_\text{EP}$ 
              corrected by hydrodynamic wall and mapping effects.
              \label{Tab:ValidElectrophoresis} }
   \begin{tabular}[h]{l|ccccc}
      \toprule
      $R_L$                                                                   &   4       &   6       &   8       &   9       &   12      \\
      \midrule 
      $U_{\text{EP},L} / \num{e-3}$                                           & 4.227     & 4.249     & 4.274     & 4.286     & 4.320     \\
      \midrule
      $U_L^* / \num{e-3}$                                                     & 4.119  & 4.144  & 4.120  & 4.135  & 4.151     \\
      $\delta_U / \si{\percent}$                                              & 2.52   & 1.52   & 0.83   & 0.68   & 0.29      \\
      $\Delta_\text{r}U / \si{\percent}$                                      & -2.56  & -2.48  & -3.59  & -3.51  & -3.91     \\
      $\Delta_\text{r}U_\text{EP} / \si{\percent}$                            & 0.5    & 0.2    & -0.6   & -0.6   & -0.9      \\
      \bottomrule
   \end{tabular}
\end{table}
As expected, the fluctuations decrease with increasing sphere resolution.
Moreover, the relative deviation of the obtained velocity $U^*$ from the theoretical electrophoretic velocity ${U_\text{EP}}$
given by $\Delta_\text{r}U = (U^*-U_\text{EP})/{U_\text{EP}}$ is listed in \Tab{Tab:ValidElectrophoresis}.
For all examined sphere radii, 
the obtained velocities in the confined domain are by $2.5\%$ to $3.9\%$ lower than the theoretical values of $U_\text{EP}$ for a particle in an unbounded electrolyte solution.
The effect of the confinement on the particle velocity is deducted
by subtracting the relative deviation from Stokes velocity $\Delta_\text{r}U_\text{Stokes}$ in \Tab{Tab:ElectrophoresisValidSpheres_DomainSizes} for the corresponding domain sizes
from the relative electrophoretic velocity deviation $\Delta_\text{r}U$ obtained in the electrophoresis simulation. 
From the resulting relative deviations ${\Delta_\text{r}U_\text{EP} = \Delta_\text{r}U - \Delta_\text{r}U_\text{Stokes}}$, 
the inaccuracies due to electric effects in the electrophoresis simulation are assessed.
As can be seen from the values of $\Delta_\text{r}U_\text{EP}$ in \Tab{Tab:ValidElectrophoresis}, 
the simulations results agree with the theoretical values with relative deviations below $1\%$.

\subsection{Parallel Performance and Scaling \label{SubSec:ParallelPerf}}
Following common practice in parallel computing, scaling experiments \cite{hager2010introduction} were performed to assess the parallel efficiency and scalability of the algorithm.
The performance experiments presented in the following were conducted on SuperMUC Phase 1.
Phase 1 comprises 18 Thin Islands with 512 Thin nodes each, connected by a high speed InfiniBand FDR10 interconnect.
Each node contains two Intel Xeon E5-2680 ``Sandy Bridge-EP`` octa-core processors that were running at 2.5 GHz, and 32 GB DDR3 RAM.
The code is built with the GCC 5.4.0 compiler, IBM MPI 1.4, and Boost 1.57.

For the performance measurements, the parameters described in~\Sect{SubSec:SimSetup} are used.
A block of size $144^3$ cells with $64$ spheres of radius $R=6 \, \textup{dx}$ is assigned to each process.
The spheres are initially arranged as a set of $4^3$ equidistantly placed particles with center-to-center distance $36 \, \textup{dx}$ and therefore weakly overlapping double layers.
Here the center of the first particle is placed at position $x = y = z = 19 \, \textup{dx}$ w.r.t\ the coordinate system in \Fig{fig:WallEffectSetup}.\\
In all experiments 240 time steps are performed and the simulation domain is periodic in y-direction.
At all other walls, no-slip conditions are applied for the LBM and homogeneous Neumann conditions for the electric potential.

For the single-node weak scaling performance evaluation, an increasing number of processes is allocated to the processors until all 16 cores are occupied.
Starting from a serial process, the problem size is extended successively in y-direction to $1\times4\times1$ processes
and then from $1\times3\times2$ to $1\times7\times2$ processes.
The maximum problem size per node with $2\times4\times2$ processes is kept constant for the weak scaling experiments.
In these experiments the problem size is extended successively in all three dimensions in a round-robin fashion.
The extension begins in z-direction with $2\times4\times4$ processes on two nodes, up to $32\times64\times32$ processes on \num{4096} nodes.
This maximum problem size comprises  $\num{196e9}$ lattice cells and $\num{4.19e6}$ particles.

For the Debye-H{\"u}ckel equation the $L_2$ norm of the initial residual has a value of \num{3.37e-3} at all problem sizes of the weak scaling experiment.
In order to permanently reduce this residual by a factor of ${\text{RES}_\text{RF} = \num{4e-6}}$, 
an average number of 29 SOR iterations per time step is required on 128 nodes and above, as observed for 640 time steps.
Therefore at each time step 29 SOR iterations are applied to solve the Debye-H{\"u}ckel equation.

About \SI{1.95}{\percent} of the domain is covered by moving obstacle cells that are not updated by the LBM and the SOR.
To quantify the performance, we use the measure \emph{million fluid lattice updates per second} (MFLUPS)~\cite{wellein2006single} 
as the number of fluid cells that can be updated within one second by the LBM and the SOR.
For the SOR this metric indicates the performance to obtain the solution up to a given accuracy, with several iterations per update.\\

\begin{figure}[h]
   \centering
     \includegraphics[width=0.6\textwidth]{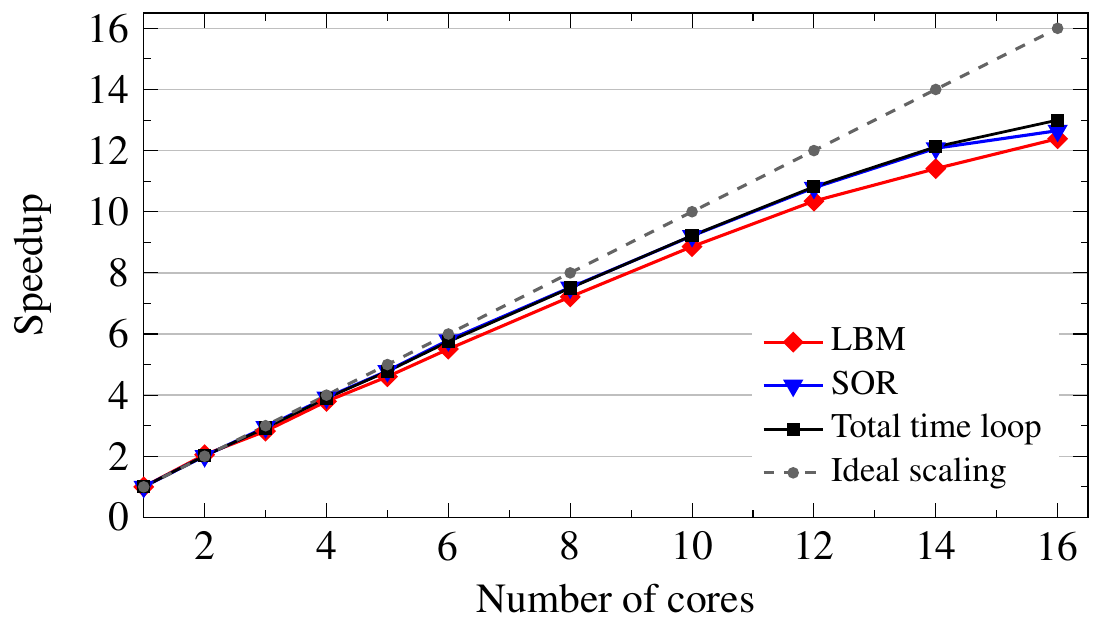}
     \captionsetup{width=0.8\textwidth}
     \caption{ Single-node weak scaling performance of the whole algorithm and SOR and LBM sweeps for 240 time steps on SuperMUC.
             }
   \label{fig:SingleNodeWeakScalSMC}
 \end{figure}

On a single node the total runtime increases from \SI{582}{\second} on one core to \SI{717}{\second} on 16 cores. 
This is equivalent to \SI{81}{\percent} parallel efficiency of the overall algorithm.
In \Fig{fig:SingleNodeWeakScalSMC} the speed\-up of the whole algorithm and its main parts is presented for up to 16 MPI processes.
Both LBM and SOR exhibit good scaling and achieve \SI{77}{\percent} and \SI{79}{\percent} parallel efficiency on a full node, respectively.
Their performance mainly restricted by computation-related data transfers, with less than \SI{2.6}{\percent} communication share between the processes for the LBM.
The relative share of intra-node communication on the SOR runtime more than doubles from \SI{3.6}{\percent} on 14 processes to \SI{7.6}{\percent} on 16 processes,
due to the additional communication in the third dimension.
The single-node performance of the LBM and the SOR sweep is 61.2 MFLUPS and 28.9 MFLUPS, respectively.
These values will serve as base performance in the following, where we employ 16 processes per node because this leads to the highest overall performance.

\begin{figure}[h]
    \centering
     \includegraphics[width=0.65\textwidth]{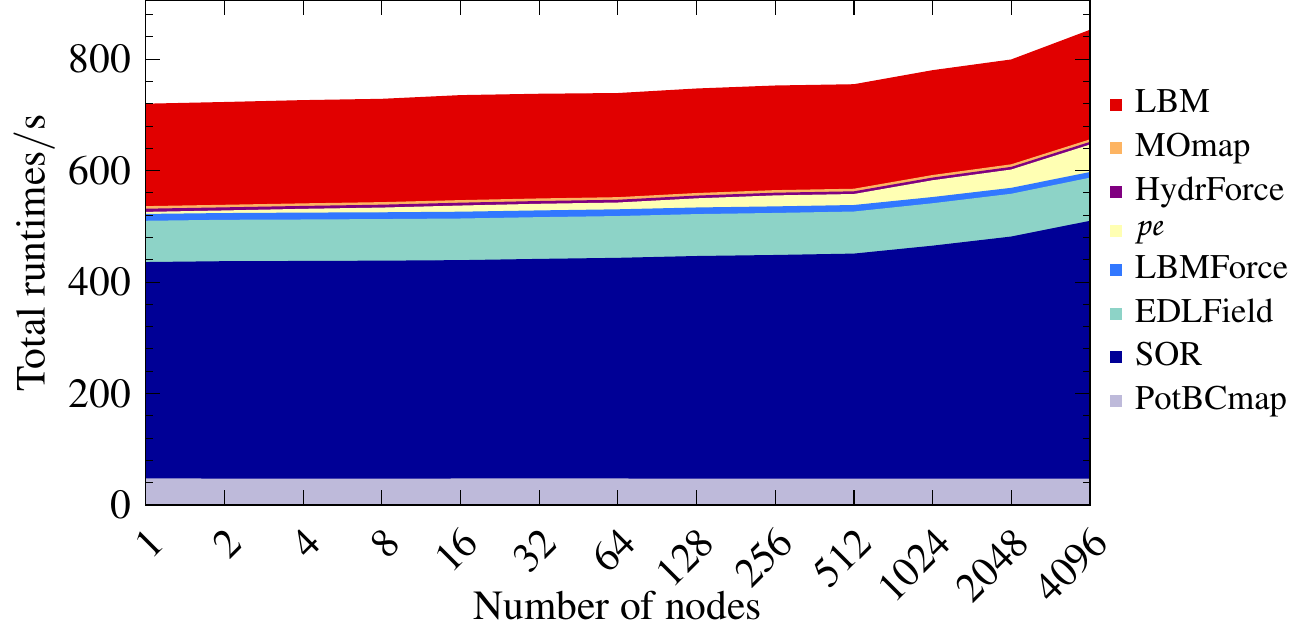}
     \captionsetup{width=0.82\textwidth}
     \caption{Runtimes of electrophoresis algorithm sweeps for 240 time steps for an increasing number of nodes (16 cores per node).}
       \label{fig:WeakScalSharesOverallAlg_240TS_SMC}
\end{figure}

In the weak scaling experiments on up to 4096 nodes on SuperMUC presented in the following for 240 time steps, 
the problem size is successively doubled as described above.
The runtimes of all parts of the algorithm are shown in \Fig{fig:WeakScalSharesOverallAlg_240TS_SMC} for different problem sizes,
indicating their shares on the total runtime.
This diagram is based on the 
maximal (for SOR, \pe{}) or average (others) runtimes of the different sweeps among all processes.
The upper part of the diagram shows the cost of fluid-simulation related sweeps, such as LBM, moving obstacle mapping (MOmap), 
and hydrodynamic force computation (HydrForce) sweep.
In the middle, the cost of the \pe{} sweep is shown.
Below, the costs of sweeps related to electric effects are displayed.
These include computing the electrostatic forces on particles and fluid (LBMForce), computing the electric field in the EDL (EDLField), 
SOR, and the sweep setting the electric potential BCs at the particles (PotBCmap).

For a more precise evaluation, the exact figures are shown in \Tab{Tab:WeakScalTimeShare} for one node and 4096 nodes.
The total runtime (Whl) differs from the sum of the individual sweeps, since different sweeps are slow on different processes.
\begin{table}[h!]
 \centering
 \caption{Time of the whole algorithm and its sweeps for 240 time steps on a single node and on 4096 nodes. Relative share of sweeps on total runtime given in parentheses.  \label{Tab:WeakScalTimeShare} }
 \begin{tabular}[h]{r|c|ccccc}
  \toprule
               & Whl    & SOR      & LBM      & \pe{} & EDLField &  Oth \\
\#n.           & t$/\si{\second}$  & t$/\si{\second}$ (\si{\percent}) & t$/\si{\second}$ (\si{\percent}) & t$/\si{\second}$ (\si{\percent}) & t$/\si{\second}$ (\si{\percent})  & t$/\si{\second}$ \\
\hline
    1          & 717   & 390 (54)  & 184 (26) & 4\,(1)   &  74 (10)  & 69  \\
    4096       & 825   & 474 (56)  & 197 (24) & 49 (6)   &  77  (9)  & 66  \\
   \bottomrule
 \end{tabular} 
\end{table}
Sweeps whose runtimes depend on the problem size---mainly due to increasing MPI communication---are LBM, MG, \pe{}, and EDLField.
Overall, LBM and SOR take up more than \SI{80}{\percent} of the total time, w.r.t.\ the runtimes of the individual sweeps.
The sweeps that scale perfectly---MOmap, HydrForce, LBMForce, and PotBCmap---are summarized as `Oth`.
Here PotBCmap has with \SI{6}{\percent} the highest portion of the total runtime.

Overall, the multiphysics algorithm achieves \SI{87}{\percent} parallel efficiency on 4096 nodes.
\begin{figure}[h]
   \centering
       \includegraphics[width=0.65\textwidth]{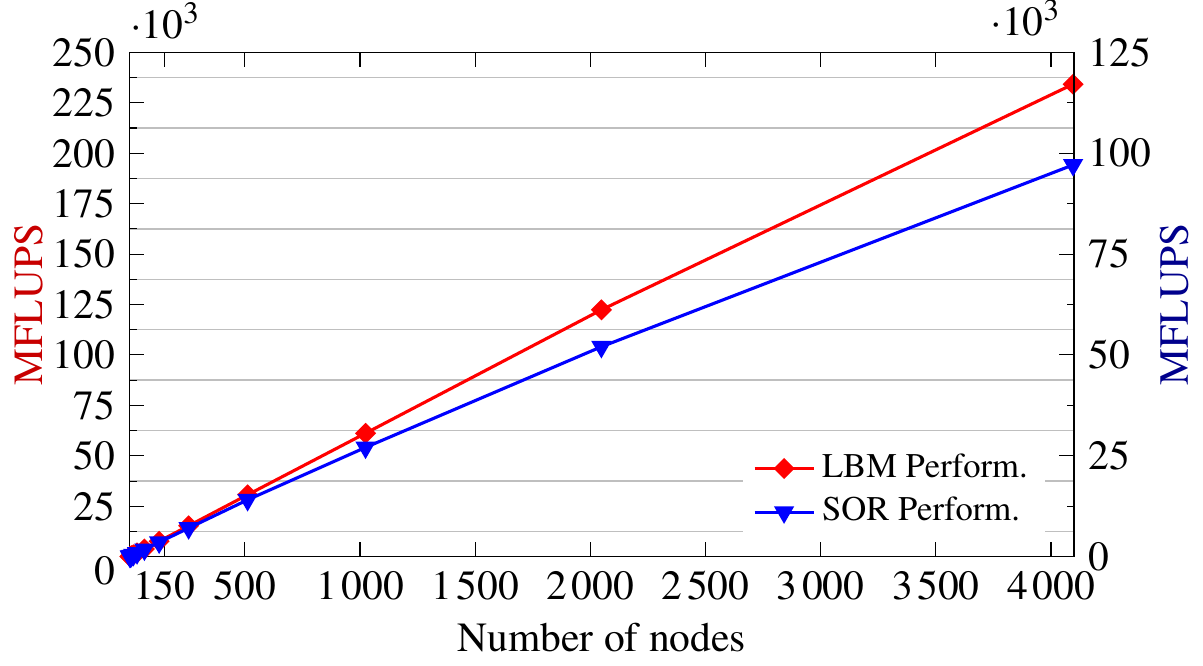}
     \captionsetup{width=0.8\textwidth}
         \caption{ Weak scaling performance of LBM and SOR sweep for 240 time steps on SuperMUC.
                   \label{fig:WeakScalMLUPs240TS_SMC}
                 }
 \end{figure}
Since most time is spent to execute LBM and SOR, we will now analyze them in more detail.
\Fig{fig:WeakScalMLUPs240TS_SMC} displays the parallel performance for different numbers of nodes.
On 4096 nodes, SOR executes \num{97102}~M\-FLUPS, corresponding to a parallel efficiency of \SI{82}{\percent}.
The LBM performs \num{234227}~M\-FLUPS, with 94\% parallel efficiency.

\begin{figure}[h!]
    \centering
     \includegraphics[width=0.69\textwidth]{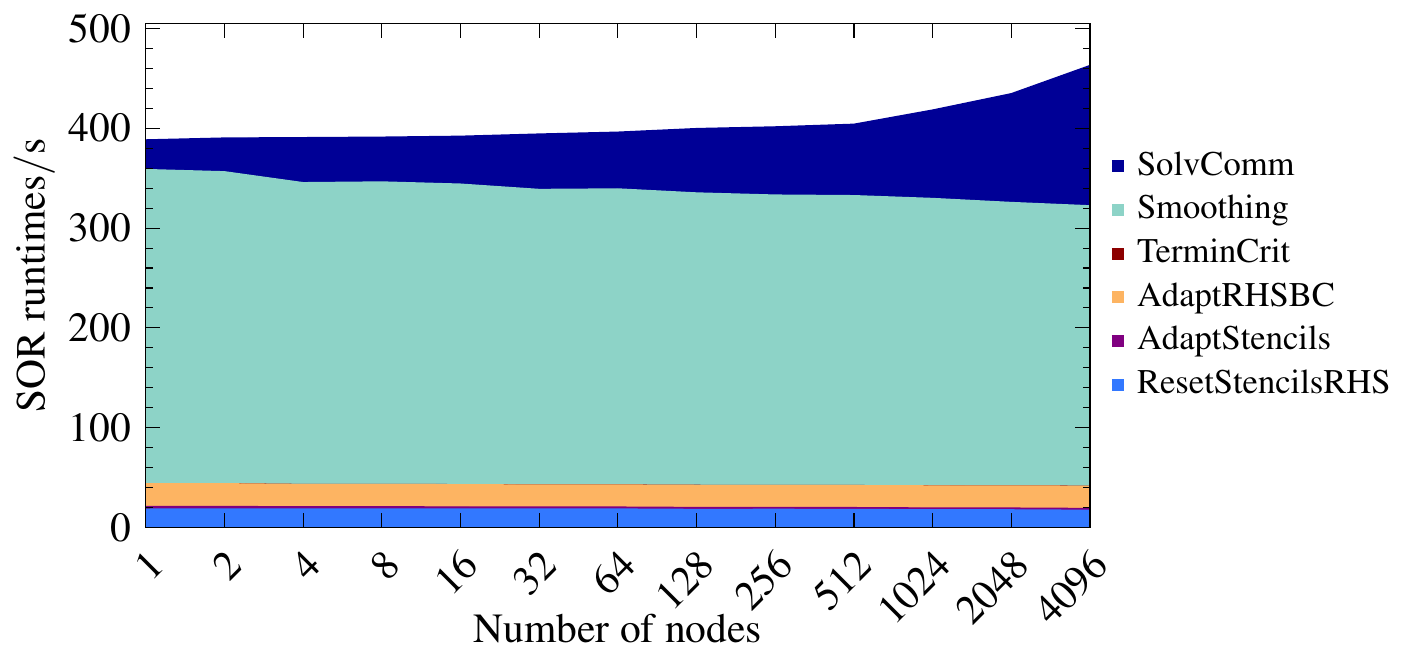}
     \captionsetup{width=0.8\textwidth}
     \caption{Average runtimes of SOR sweeps for 240 time steps with 29 iterations per time step.}
   \label{fig:WeakScalSharesSOR_240TS_SMC}
 \end{figure}

Finally, the average execution times of the different SOR sub-sweeps are presented in \Fig{fig:WeakScalSharesSOR_240TS_SMC}.
Like in the single-node scaling, the parallel performance of the SOR is degraded mainly by increasing communication costs.
All other sweeps exhibit ideal scaling. 
Among these sweeps, most time is spent for smoothing, whereas all other parts require only a small portion of the overall SOR time.
These are the sweeps for resetting the stencils and right-hand side to their initial values (ResetStencilsRHS, \SI{18}{\second}), 
for adapting the right-hand side to the BCs (AdaptRHSBC, \SI{22}{\second}), and adapting the stencils to the BCs (AdaptStencils, \SI{2}{\second}).
The sweep for checking the termination criterion is negligible ($<\SI{0.02}{\second}$).
\section{Conclusion \label{Sec:Conclusion}}
In this article, a coupled multiphysics algorithm for parallel simulations of the electrophoretic motion of geometrically resolved particles in electrolyte solutions is presented.
The physical effects are simulated by means of the lattice Boltzmann method for fluid dynamics,
a physics engine for rigid body dynamics, and a scalar iterative solver for electric potentials.
These components are integrated into the parallel software framework \walberla
to simulate the electrophoretic motion of fully resolved charged particles in microfluidic flow.
The simulations include fluid-particle and electrostatic particle interactions.
Additionally electric effects on ions around the homogeneously charged particles are recovered.

The current work is an extension of \cite{Bartuschat:2014:CP}, where the electrical migration of charged particles without ions in the fluid was validated and excellent parallel performance was shown for more than seven millions of interacting charged particles.
In the present article, the opposite net charge in the electric double layer (EDL) around the charged particles due to ions in the fluid is considered, together with its effect on the fluid motion that 
counteracts particle motion.
To this end, the electric potential distribution in the fluid due to the EDLs is computed that causes an electric body force on the fluid. 
This quasi-equilibrium distribution recovers the motion of ions in the fluid along with the charged particles while neglecting EDL distortion.

The overall electrophoresis algorithm is introduced and an overview of the coupled functionality implemented in the involved \walberla modules is given.
For the simulations, a solver sweep for time-varying boundary conditions has been developed that is presented here for the parallel SOR method employed to solve the EDL potential equation.
Based on the multiphysics boundary handling concept \cite{Bartuschat:2014:CP} an efficient parallel algorithm is implemented to impose electric potential boundary conditions on the moving particles.
These methods can also be employed for other governing equations with spatially varying boundary conditions that model physical effects different from electric fields.
The presented parallel electrophoresis simulations are also facilitated by a joint parameterization concept for the different coupled governing equations
and numerical methods implemented in \walberla. This concept is based on lattice Boltzmann requirements and is applicable and extensible to further multiphysics simulations.

For the electrophoresis simulations in this article, the electric potential in the double layer is shown to coincide with analytical solutions.
The obtained terminal electrophoretic velocities comply with analytical solutions for different proportions of the particle radii to double layer thickness.
These validation results verify the correctness of the implementation and the coupling of the different methods.
Moreover, the observed relative errors in the modeling of electric effects are below \SI{1}{\percent}.
The retardation effect caused by the presence of the EDL is shown to be significant for the examined sphere radii, reducing the sphere velocity up to \SI{42}{\percent}.
For the electrophoretic motion in a micro-channel, the flow field and the electric potential distribution are visualized,
including the ion charge distribution in the EDL surrounding the particle.

The parallel performance of the algorithm is examined on SuperMUC. 
The weak scaling of the overall algorithm yields a parallel efficiency of \SI{87}{\percent} on \num{65536} cores, simulating more than four million charged particles and their surrounding by EDLs with \SI{1.95}{\percent} solid volume fraction.
The overall parallel performance is based on a carefully designed, extensible software architecture that exceeds the capabilities of alternative simulation software by several orders of magnitude.
Most time of the electrophoresis algorithm is spent for the SOR solver and the LBM.
The SOR algorithm scales with \SI{82}{\percent} parallel efficiency, achieving more than \num{97.1e9} fluid cell updates per second to solve the Debye-H\"uckel approximation. 
The LBM for fluid-particle interaction scales almost perfectly with \SI{94}{\percent} parallel efficiency, achieving more than \num{234e9} fluid cell updates per second.

The presented algorithm can be applied to find design parameters in industrial and medical applications,
\eg, for optimal separation efficiency of charged biological particles in lab-on-a-chip devices, depending on fluid, particle, and electrolyte properties.
Our algorithms were shown to correctly recover fluid-particle interactions for elongated particles in \cite{Bartuschat:2014:Tumbling}.
These future simulations may therefore include suspended, possibly charged particles of various shapes
including spheres, spherocylinders, and particles of more complex shapes,
\eg, to represent different biological particles.
Also pairwise van der Waals forces can be added easily, to facilitate simulations of electrophoretic deposition in material science applications.

The electrophoresis algorithm introduced here is well suited for massively parallel simulations.
In the current implementation of this algorithm, the EDL thickness is restricted to values in the order of the particle radius.
Therefore, adaptive lattice refinement as in \cite{schornbaum2016massively} may be employed to allow for thinner double layers relative to the particle size.
For the incorporation of transient effects in the simulations including EDLs,
the link-flux method implemented into \walberla in \cite{hufnagelBth} and \cite{kuron2016moving} may be employed.
This method was extended in \cite{kuron2016moving} to simulate electrophoresis, enabling the simulation of non-equilibrium ion distributions in the EDL.
Due to the higher computational complexity of the link-flux method compared to the equilibrium approach in this article,
the maximum number of particles will be lower than in our approach.%

\section*{Acknowledgements}
\noindent
The authors are grateful to the LRZ for providing the computational resources on SuperMUC.




\bibliographystyle{elsarticle-num} 


\end{document}